\documentclass[preprint,5p,twocolumn,times]{elsarticle}%

\usepackage{xparse}%
\usepackage{morewrites}
\usepackage{wrapfig}
\usepackage{xkeyval}%
\usepackage{verbatim}

\usepackage{amsmath}
\usepackage{algorithm}
\usepackage{algpseudocode}
\makeatletter
\def\BState{\State\hskip-\ALG@thistlm}
\makeatother

\newwrite\authorbibfile%

\AtBeginDocument{%
  \immediate\openout\authorbibfile=\jobname.aub%
}%

\AtEndDocument{%
\immediate\closeout\authorbibfile
\InputIfFileExists{\jobname.aub}{}{}
}%

\makeatletter

\define@key{authorbib}{scale}[1]{%
\def\AuthorbibKVMacroScale{#1}%
}

\define@key{authorbib}{wraplines}[10]{%
\def\AuthorbibKVMacroWraplines{#1}%
}

\define@key{authorbib}{imagewidth}[4cm]{%
\def\AuthorbibKVMacroImagewidth{#1}%
}

\define@key{authorbib}{overhang}[10pt]{%
\def\AuthorbibKVMacroOverhang{#1}%
}

\define@key{authorbib}{imagepos}[l]{%
\def\AuthorbibKVMacroImagepos{#1}%
}

\makeatother

\presetkeys{authorbib}{imagepos=l,imagewidth=4cm,wraplines=15,overhang=20pt}{}

\newlength{\AuthorbibTopSkip}
\newlength{\AuthorbibBottomSkip}
\NewDocumentCommand{\authorbibliography}{+o+m+m+m}{%
  \IfNoValueTF{#1}{%
  }{%
    \setkeys{authorbib}{#1}%
    \immediate\write\authorbibfile{%
      \string\begin{wrapfigure}[\AuthorbibKVMacroWraplines]{\AuthorbibKVMacroImagepos}[\AuthorbibKVMacroOverhang]{\AuthorbibKVMacroImagewidth}^^J
        \string\includegraphics[scale=\AuthorbibKVMacroScale]{#2}^^J
        \string\end{wrapfigure}^^J
    }%
  }%
  \IfNoValueTF{#3}{%
    \typeout{Warning: No author name}%
  }{%
    \immediate\write\authorbibfile{%
      \unexpanded{\vspace{\AuthorbibTopSkip}}^^J
      \string\noindent\relax
      \unexpanded{\textbf{#3}\par}^^J
      \string\noindent\relax
      \unexpanded{#4}^^J%
      \unexpanded{\vspace{\AuthorbibBottomSkip}}^^J
      }%
  }%
}%

\usepackage{amssymb}
\usepackage{booktabs} %
\usepackage{subcaption}
\usepackage{algorithm}
\usepackage{algpseudocode}

\usepackage{url}

\usepackage[breaklinks,colorlinks=true,linkcolor=black,hidelinks]{hyperref}
\newcommand{\MYhref}[3][blue]{\href{#2}{\color{#1}{#3}}}

\usepackage{cmap}				%
\usepackage{xcolor}
\usepackage{pdfpages}
\usepackage{bibentry}       %
\usepackage{fancyvrb}       %
\usepackage{makecell}       %
\usepackage{graphicx}       %
\usepackage{multirow}

\usepackage[capitalise, nameinlink, noabbrev]{cleveref}

\DefineVerbatimEnvironment{Code}{BVerbatim}{baseline=t}

\usepackage{soul}
\usepackage{ulem}

\begin{document}

\begin{frontmatter}

\title{TunaOil: A Tuning Algorithm Strategy for Reservoir Simulation Workloads} %

\author[Petrobras,BSC,UPC]{Felipe Portella\corref{cor1}}
\ead{felipe@portella.com.br}
\author[BSC,UPC]{David Buchaca}
\ead{david.buchaca@bsc.es}
\author[Petrobras]{Jos\'e Roberto Rodrigues}
\ead{jrprodrigues@petrobras.com.br}
\author[BSC,UPC]{Josep Ll. Berral}
\ead{josep.berral@bsc.es}

\address[Petrobras]{Petr\'oleo Brasileiro S.A. (PETROBRAS),
Av. Hor\'acio de Macedo, 950, 21.941-915, Rio de Janeiro, Brazil}
\address[BSC]{Barcelona Supercomputing Center (BSC),
C. Jordi Girona 1-3, 08034, Barcelona, Spain}
\address[UPC]{Universitat Polit\`{e}cnica de Catalunya (UPC) - BarcelonaTech,
Campus Nord, Carrer de Jordi Girona, 1, 3, 08034 Barcelona, Spain}

\cortext[cor1]{Corresponding author}

\begin{abstract}

Reservoir simulations for petroleum fields and seismic imaging are known as the most demanding workloads for high-performance computing (HPC) in the oil and gas (O\&G) industry. The optimization of the simulator numerical parameters plays a vital role as it could save considerable computational efforts. State-of-the-art optimization techniques are based on running numerous simulations, specific for that purpose, to find good parameter candidates. However, using such an approach is highly costly in terms of time and computing resources. 
This work presents TunaOil, a new methodology to enhance the search for optimal numerical parameters of reservoir flow simulations using a performance model. 
In the O\&G industry, it is common to use ensembles of models in different workflows to reduce the uncertainty associated with forecasting O\&G production. We leverage the runs of those ensembles in such workflows to extract information from each simulation and optimize the numerical parameters in their subsequent runs.

To validate the methodology, we implemented it in a history matching (HM) process that uses a Kalman filter algorithm to adjust an ensemble of reservoir models to match the observed data from the real field. We mine past execution logs from many simulations with different numerical configurations and build a machine learning model based on extracted features from the data. These features include properties of the reservoir models themselves, such as the number of active cells, to statistics of the simulation's behavior, such as the number of iterations of the linear solver. A sampling technique is used to query the oracle to find the numerical parameters that can reduce the elapsed time without significantly impacting the quality of the results. 
Our experiments show that the predictions can improve the overall HM workflow runtime on average by 31\%.

\end{abstract}

\begin{keyword}
Petroleum Reservoirs \sep Reservoir Simulation \sep Parameter Tuning \sep Machine Learning \sep Performance Model 
\end{keyword}

\end{frontmatter}

\section{Introduction}\label{sec:introduction}

Reservoir simulation using computer modeling is a fundamental tool in the field of petroleum reservoir engineering. Engineers perform numerical flow simulations to decide how to exploit a petroleum field, and these simulations can be costly in terms of execution time. Tuning its numerical parameters is a standard procedure that can significantly reduce that time, although it might impact the quality of the results. Reservoir simulators contain a huge amount of software tunable parameters that affect the runtime of the application and the quality of the predictions made by the reservoir model. Tuning is complex as the goal is to find a set of parameters for the simulation that minimizes the simulation execution time, maintaining good quality results. We can cast this challenge as an optimization problem with a quality constraint.

Engineers can manually tune the numerical parameters of the simulation. However, this is challenging because it requires extensive knowledge about the simulator. Moreover, it is hard to manually balance many competing objectives, such as the execution time and the quality of the simulation (e.g., material balance error). Optimization programs are widespread in the reservoir simulation field, and engineers typically use academic or commercial tools to make strategic and operational decisions. These tools can be used to tune the numerical parameters, having set only a search space and an objective function. 
The downside of a ``traditional'' optimization process, such as a Genetic Algorithms\cite{Lima2015} or a Bayesian Optimization\cite{Snoek2012}, is its execution cost. Dozens to hundreds of simulations may be required to find decent solutions. 
This paper proposes an approach to tune these numerical settings automatically while reducing the overall computational runtime. 
Our solution - TunaOil - has the benefit of optimizing the numerical parameters without the need for a separate study and additional simulations. 
To achieve that, we embedded the tuning process into a standard reservoir engineering process known as history matching (HM), a data assimilation process which adjusts the fluid flow simulation model to reproduce the past behavior observed in the real-life reservoir field. Several HM techniques are available, but methods based on Kalman filters stand out due to several advantages, including the capability of quantifying uncertainty through adjustments to multiple models at the same time.
In this work, we use one of these methods -- the Ensemble Smoother with Multiple Data Assimilation (ES-MDA)~\cite{Emerick2013} -- to incorporate the proposed optimization methodology that embeds a performance model. However, any other process requiring multiple runs of the reservoir models could apply our methodology, such as optimizing the well-placement or conducting a sensitivity analysis of some rock-fluid parameters.

The proposed methodology uses the first round of simulations of the ES-MDA to extract from the logs a feature vector of the reservoir model and the current performance, refit the regression, and then is able to suggest new numerical settings for further executions. We evaluated the method on different public reservoir models with multiple realizations using an HM workflow based on the ES-MDA. Our new methodology reduces the overall processes execution without a substantial negative quality impact in the reservoir simulation. There is a trade-off between the elapsed time and the quality, but the quality of the flow simulation does not need to exceed a certain threshold. In other words, if it is ``good enough'', there is no practical need to make it better at the expense of increasing the run time of the simulator.

Reservoir simulations are one of the most demanding High-Performance Computing (HPC) workloads in the industry \cite{Mustafiz2008}. Any slight improvement in the execution time of this application represents significant savings for Energy companies that depend on these simulations for their very different daily-base operations. The monetary costs are substantial as those companies use expensive supercomputers, 
many of which are on the TOP500\footnote{\MYhref{https://www.top500.org/statistics/sublist/}{https://www.top500.org/statistics/sublist/}} list. Moreover, engineers can explore more scenarios and make quicker decisions with faster simulations. 

TunaOil achieves a faster overall runtime of the HM process by dynamically tuning the simulations executed in the HM workflow. Our experiments show that TunaOil can improve the execution time of the overall workflow by at least 23\% while increasing the quality metric (the material balance error) by less than 1\%. Moreover, it does not require simulations specific to the purpose of numerical tuning. 

\section*{Contributions}

The main contributions of our work can be summarized as follows:

\begin{itemize}

    \item 
    A novel methodology leveraging previous reservoir simulation executions to model and propose near-optimal numerical parameters for following simulations, using a Machine Learning (ML) based strategy embedded into the HM method. Such a strategy allows the simulation parameters to be adjusted without running additional executions, thus saving execution time.
    
    \item A method for using extracted information from reservoir simulation logs, providing a novel feature vector with critical knowledge for training performance models.
    
    \item The evaluation of three reservoir studies through our methodology, demonstrating it reduced in the HM workflow's overall execution time.

\end{itemize}

\section*{Paper structure}

The rest of this paper is structured as follows: Section~\ref{sec:background} provides background on Reservoir Simulation, History Matching, and the ES-MDA. It also contains some discussion of related works in tuning the numerical parameters. Section~\ref{sec:problem} details the problem being solved. Then, Section~\ref{sec:proposal} describes the proposed workflow, its implementation within the ES-MDA, and the performance model. Section~\ref{sec:workloads} briefly presents the workloads that were used in the experiments of Section~\ref{sec:experiments}, while Section~\ref{sec:conclusion} concludes the paper.

\section{Background}\label{sec:background}

This section gives an overview of the process of petroleum reservoir simulation. This includes a description of reservoir simulation models, History Matching (HM), and how Kalman filters are applied to do the HM, as well as a specific Kalman technique called Ensemble Smoother with Multiple Data Assimilation (ES-MDA) that is used in our work to demonstrate the advantages of our proposal. We also present the related works on techniques for numerical tuning of flow simulation.

\subsection{Reservoir Simulation}

Reservoir simulators are extensively used in the petroleum industry as they allow engineers to forecast oil, gas and water production from the reservoir providing support to business decisions regarding the exploitation of the oilfield. By simulating the complex fluid flow processes in different exploitation alternatives for the petroleum field, engineers can answer a series of critical questions for field development and management, such as the number and location of wells to be drilled and the best recovery method to be used, among others.

Reservoir simulation models are mathematical models that describe the flow in a porous medium. Such models are based on conservation of mass and energy equations, together with some empirical relationships, e.g., Darcy's law. In general, these equations cannot be solved analytically, and numerical techniques must be applied to find their solutions. 
One of the most popular numerical methods currently used in reservoir engineering is the finite difference (FD) method, employing two-point flux approximation (TPFA)\cite{Aziz1979} to discretize Darcy's law.
The FD discretization results in a nonlinear system of algebraic equations in each time step to determine pressure and saturations (or compositions) in each mesh cell. In modern simulators, some variant of Newton's method is used, requiring the solution of a linear system at each iteration. Due to its complexity and huge size, iterative methods are the only practical choice to solve those linear systems. 
Works \cite{peaceman1977} and \cite{Aziz1979} provides an in-depth look into the methods used in reservoir simulation and work \cite{Collins2003} gives details about the parallel ILU Solver used in the experiments.

Time step control, and the iterative nature of linear and nonlinear solution methods, require several numerical parameters to be defined, such as minimum and maximum time step sizes, convergence tolerances, and the maximum number of iterations, to name just a few. Engineers are constantly looking for efficient tools to tune the simulation process to make it faster, fulfilling the continuous business pressure to make better decisions. The reservoir simulators available on the market allow users to tune those numerical parameters, which can significantly affect the performance and quality of the simulation. However, these numerical parameters vary between simulators, making their selection difficult. Moreover, the parameter space is big, and co-relations can exist between parameters, making this a non-trivial manual task.

The perceived computational performance of any reservoir simulator is the elapsed (or wall) time it takes to produce the prediction from a reservoir model for a specified period (e.g., simulating the 10-year history and forecasting 20 years into the future). The elapsed time of a reservoir simulation depends mainly on the input data, the numerical controls of convergence, and the efficiency of its internal solvers. The input data, which defines the reservoir model itself, has several factors impacting simulation time: the number and shape of the cells (grid data), rock properties, and fluids data, among others.

\subsection{Material Balance Error}

Since the fundamental physical law used in the simulation models is mass conservation, the Material Balance Error (MBE), which measures the amount of mass that has been ``lost'' or ``created'' due to inaccuracies in the solution process, is used as a quality control metric. In a black-oil simulator, as used in this work, the MBE accounts for three phases: oil, gas, and water. The MBE for each phase is given by:

\begin{equation}\label{eq:MBE}
MBE = \left \{ \frac{FIP}{(OFIP - Prod + Inj)} - 1 \right \} \times 100\%,
\end{equation}
where $FIP$ is the fluid in place at time $t$, $OFIP$ is the original fluid in place, $Prod$ is the cumulative fluid produced at time $t$ and $Inj$ is the cumulative fluid injected at time $t$. Based on this definition, the MBE should be equal to zero for each fluid phase present in the reservoir in the absence of numerical errors. However, a low value for the MBE is not sufficient to ensure that the numerical model is correct, since it measures the discrepancy in mass conservation for the whole reservoir, whereas the FD discretization aims to ensure material balance at each cell. The MBE is a quick way to evaluate if any changes, especially in the numerical parameters of the simulator, introduced large errors to the basic equations of the reservoir system. For practical purposes, we will be evaluating the quality of the changes in the numerical parameters using the cumulative MBE as part of our performance model. For final validation, we will also cherry-pick some models to compare the results in terms of injection and production of oil, gas, and water.

\subsection{History Matching}

Engineers use a \textit{History Matching} (HM) method to calibrate the reservoir model. HM will adjust the model's physical parameters to reproduce the past behavior of the actual reservoir field as closely as possible. It could change, for instance, the model's permeability or porosity maps to match the observed pressures and productions. Therefore, HM will reduce the uncertainty of the predictions by increasing confidence in the reproduction of the reservoir field behavior, leading to better forecasting.

History matching can be done either manually, by the reservoir engineer, or in an assisted way with the aid of computational tools guided by the professional. The manual, trial-and-error process is still used nowadays, but it takes time and requires accurate judgment from the practitioner making the adjustments. The assisted process aims to minimize a pre-defined objective function measuring the mismatch between simulated and observed data. One popular tool used for assisted HM is the family of ensemble Kalman filters.

\subsection{Ensemble Kalman Filters}

\begin{figure}[tb]
\centering
\includegraphics[scale=0.21,trim=40 180 100 180,clip]{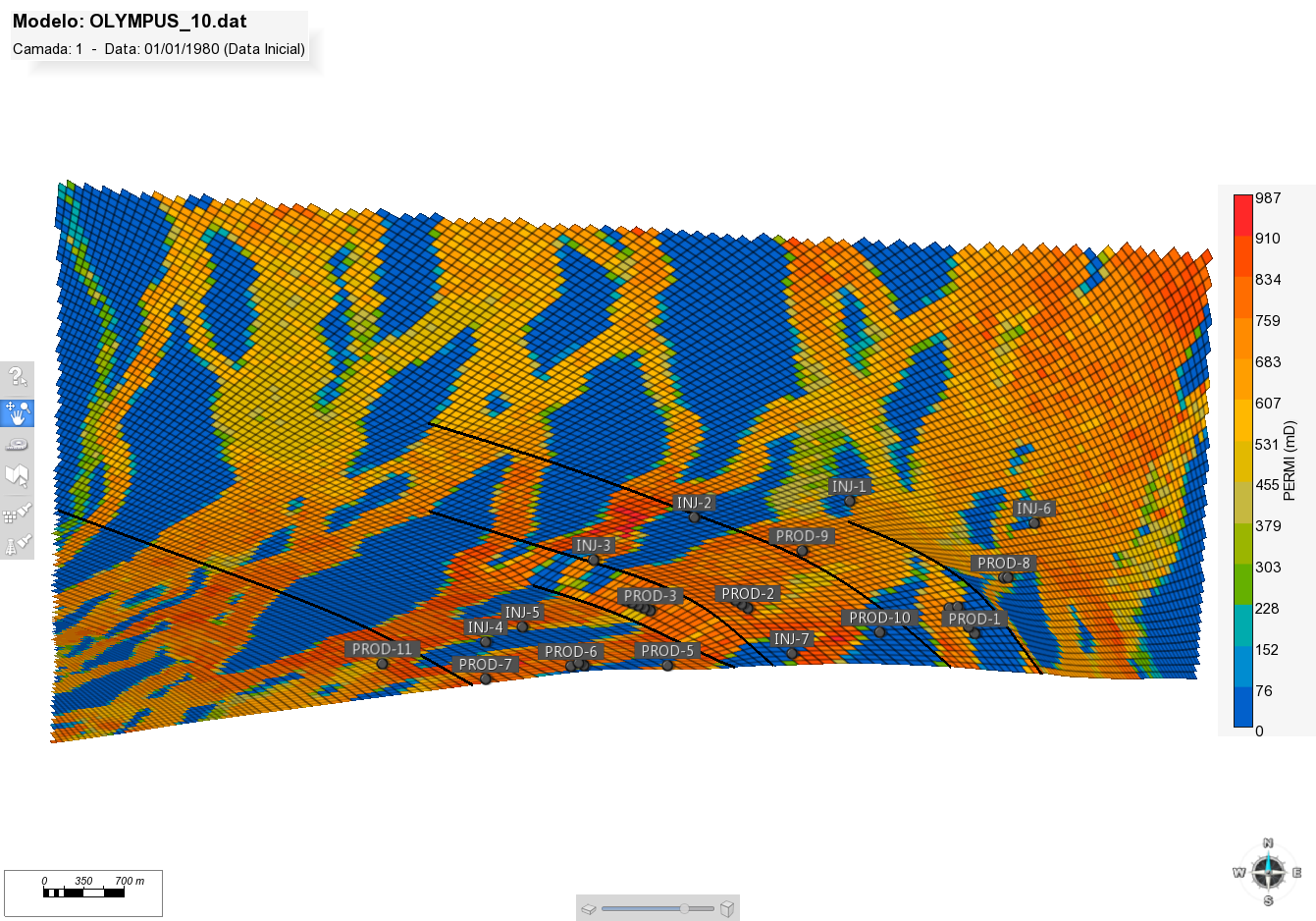}
\includegraphics[scale=0.21,trim=40 180 100 180,clip]{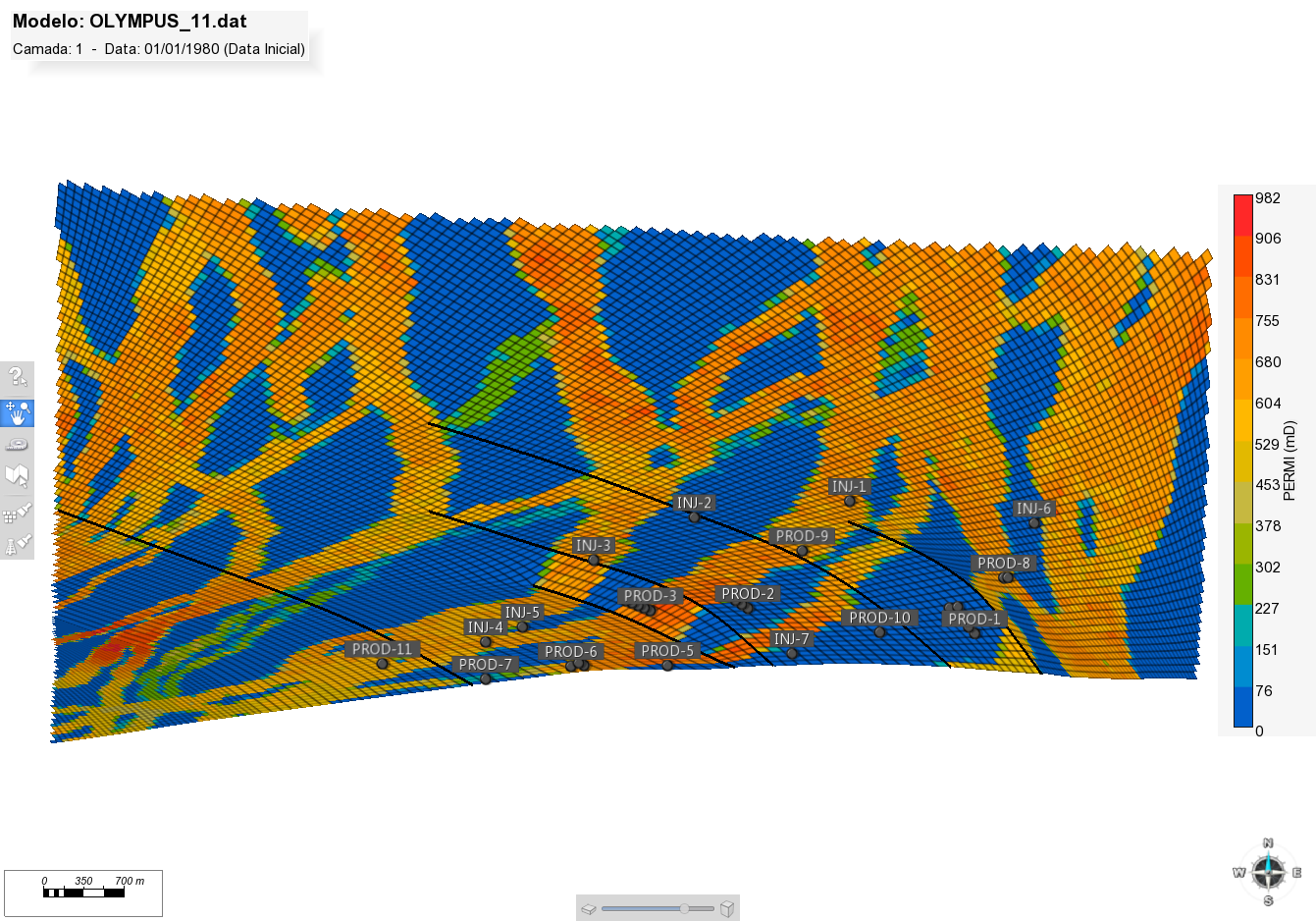}
\includegraphics[scale=0.21,trim=40 180 100 180,clip]{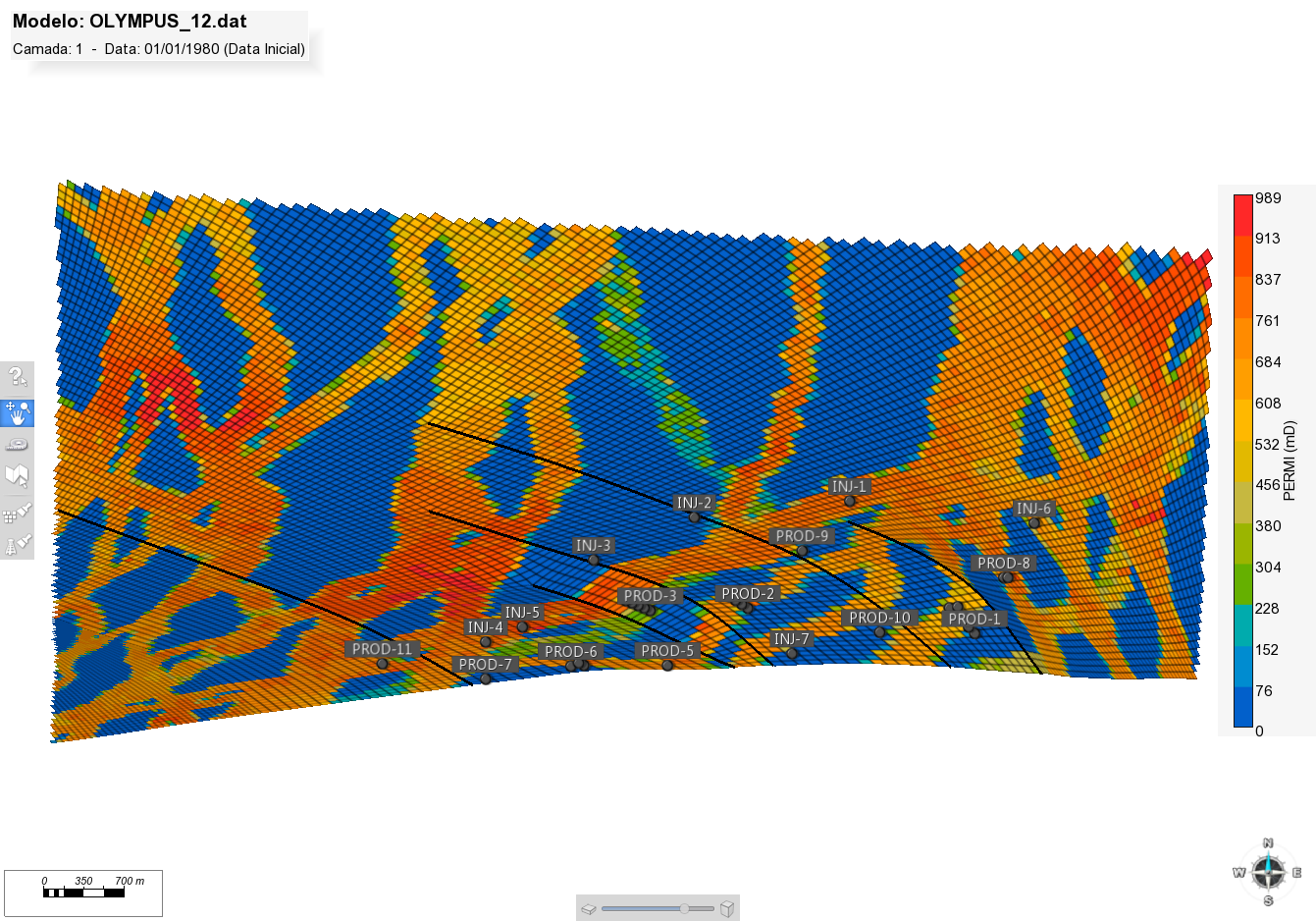}
\includegraphics[angle=-90,scale=0.35,trim=1250 180 10 180,clip]{images/OLYMPUS_12_Layer1_PERMI.png}
\caption{
Three illustrations of the permeability of the first layer with its channels, selected from the 50 model realizations available for the OLYMPUS workload.
}
\label{fig:workload_olympus}
\end{figure}

The Ensemble Kalman Filter (EnKF)~\cite{Evensen1994} is a Monte Carlo extension of the well-known Kalman filter~\cite{Kalman1960}, capable of working on non-linear systems, making it perfectly applicable to the problem of HM in reservoir simulation. This family of methods is extremely computational demanding as it uses an ensemble of models. The ensemble consists of a set of reservoir models, called realizations, with different property maps as shown in~\cref{fig:workload_olympus}. Various extensions have been proposed for the petroleum industry, such as the Ensemble Kalman Smoother (ES)~\cite{Evensen2000} and the Ensemble Kalman Filter with Multiple Data Assimilation (EnKF-MDA)~\cite{Emerick2012}.

One of the most widely used extensions is the Ensemble Smoother with Multiple Data Assimilation (ES-MDA)~\cite{Emerick2013}, which makes repeated assimilations of the same data. Work \citep{Emerick2013} reports satisfactory results with only four data assimilations plus one extra simulation to predict the forecast. The implementation is relatively simple but computationally demanding as the ensemble of models is simulated multiple times.

\subsubsection{ES-MDA}\label{sec:background:kalman:formulation}

Unlike the traditional EnKF, which performs assimilation steps during the simulation time, ES-MDA simulates from start to finish, updating the model parameters at the end. The process of updating the model parameters in the ES-MDA can be described by the following equation:
\begin{equation}\label{eq:EnKF}
M_{j}^{n+1} = M_{j}^{n} + K^{n} (d_{obs,j}^{n} - d_{sim,j}^{n}), \text{ for  } j=1,...,N_r
\end{equation}
where $M_{j}^{n}$ is the vector with model parameters (the static variables such as permeability and porosity), $K^{n}$ is the Kalman gain matrix, $d_{obs,j}^{n}$ the vector of perturbed observed data, $d_{sim,j}^{n}$ the vector of simulated data, and $N_r$ is the number of models in the ensemble.  Subscript ${j}$ is the model index in the ensemble, while superscript $n$ is the assimilation step index, $n = 0,...,N_a - 1$. A pre-defined number $N_a$ of assimilation steps determines how many times the ensemble models will be run and the models updated. 

The Kalman gain matrix $K$ is defined as follows:
\begin{equation}\label{eq:KalmanGain}
K^{n} = C_{MD}^{n} ({\alpha}^{n+1}C_{D}^{n}+C_{DD}^{n})^{-1},
\end{equation}
where $C_{MD}$ is the cross-covariance matrix between the parameter vectors and the simulated data, $C_{D}$, the error covariance matrix of the observed data (usually taken as diagonal), and $C_{DD}$, the simulated data covariance matrix. ${\alpha^{n+1}}$ are inflation factors that satisfy 
\begin{equation}\label{eq:alpha}
\sum_{n=1}^{N_a}\frac{1}{{\alpha}^{n}} = 1.
\end{equation}

The term $C_{MD}$ from Eq.(\ref{eq:KalmanGain}) can be approximated as:
\begin{equation}\label{eq:CYD}
C_{MD}^{n} = \frac{1}{N_r - 1}\sum_{j=1}^{N_r}(M_{j}^{n}-\overline{M_{j}^{n}})(d_{sim,j}^{n}-\overline{d_{sim,j}^{n}})^T,
\end{equation}
where $\overline{M_{j}^{n}}$ and $\overline{d_{sim,j}^{n}}$ are the ensemble means of the parameters and the simulated data, respectively. A similar expression is employed to calculate $C_{DD}^{n}$ with $M_{j}^{n}$ and $\overline{M_{j}^{n}}$ substituted by $d_{sim,j}^{n}$ and $\overline{d_{sim,j}^{n}}$.

The covariance given by Eq.(\ref{eq:CYD}) suffers from the size of the ensemble being relatively small in comparasion to the number of model parameters, leading to sampling errors that can result in spurious correlations \cite{Aanonsen2009}. This could cause changes in the parameters for regions that are not influenced by the assimilated data. Localization techniques \cite{Houtekamer1998,Yeo2010,Arroyo2008} restrict the ES-MDA update only to the mesh cells related to the region of the measured data and are mandatory for the application of the method in any production environment; however, to keep the implementation simple for research purposes, it was left out of the TunaOil code listed in \cref{alg:pseudocode}. Its absence will not impact the reservoir simulation numerical tuning evaluation within an HM workflow, which is the main objective of this research. In fact, as mentioned before, the proposed methodology could be applied in any workflow that uses multiple runs. It was chosen to use the HM with the ES-MDA as a simple to implement, yet relevant, workflow in the O\&G industry. 

\subsection{Numerical Tuning}\label{sec:relatedwork}

The tuning of flow simulators parameters has been widely explored in the O\&G industry using techniques mostly on classical optimization. Additionally, it can be seen in a more general way as an algorithm configuration problem. This section will present some related works exploring these methodologies and techniques.

\subsubsection{Classical Reservoir Simulation Optimization}\label{sec:background:simulation_tuning}

In~\cite{Avansi2019}, the authors apply the DECE (Designed Exploration and Controlled Evolution) \cite{Ajayi2019} method twice successively. The first performs a random level selection for each parameter with a Tabu Search\cite{Glover1989} and Experimental Design\cite{Lundstedt1998}, while the second optimizes the objective function (the elapsed time in the case).
The innovation in their approach was their methodology, which performed a numerical diagnosis structured with seven steps. They used four metrics as criteria for convergence and accuracy: the total (oil + water + gas) material balance error; the number of time-step cuts per time-step; the number of Newtonian cycles per time-step; and the number of solver iterations (SI) per Newtonian cycle. They optimized 12 numerical parameters using as baseline the default settings for the same black-oil reservoir simulator we used in our work. They evaluated the proposed methodology in two field models: META-D-h2v2 and UNISIM-I-D. For the former, they used 1,350 simulations, and for the latter 1,500. This improved performance by reducing the elapsed time by around 45\%, with an MBE increase of 0.2\% in the worst case.

Another example of innovation over the classical optimization approach is given in \cite{Rios2020}. The authors propose a workflow to select a representative numerical submodel (a slice of the overall grid) or a representative time interval (instead of the entire original period being simulated) that can be used in the optimization process and be representative of the whole model. The optimization is then performed on this smaller problem (in size or time) to tune the numerical parameters. The values found are then applied to the original full-scale reservoir numerical model. This is significantly quicker than running the optimization process with the full-scale (original) models. They perform a final consistency check step, after the numerical tuning, to verify the quality of the results using the MBE. 
They compare the performance of the numerical tuning against the default parameters of two related models: UNISIM-I-D-h7v7 (an 81x58x23 corner-point grid model) and UNISIM-I-D-hfm (a 326x234x157 corner-point grid model, almost 10x the size of the  smaller one in terms of total cells). Their search space consisted of the same 12 parameters used in work \cite{Avansi2019}.
They did not consider parallelization as one of the parameters and used only one or two processors in the evaluation. In our work we choose to use the same parameters plus another three detailed in~\cref{tab:nrs_features}.

\subsubsection{Algorithm Configuration}

The specific literature in the Algorithm Configuration (AC) field has several approaches, ranging from Direct Search Methods to Model-based optimization or Genetic Programming. Many have been used to optimize algorithm performance, and the majority evaluate each parameter configuration by combining experiment designs (full factorial or fractional factorial) and gradient descent techniques. Work~\cite{Huang2020} provides a recent survey of automatic parameter tuning methods, and the authors propose a taxonomy, with three main categories according to the structure of tuning methods, for classification of these methods for metaheuristics. These are simple generate-evaluate methods, iterative generate-evaluate methods, and high-level generate-evaluate methods.

The most straightforward approach is to generate some parameter candidates and evaluate them to find the best configurations. Brute-force methods or the F-Race~\cite{Birattari2002} falls into this simple generate-evaluate category where all the candidate configurations are generated at once and evaluated sequentially. These algorithms are simple to understand and implement but computationally expensive, particularly with large parameter spaces. In the iterative generate-evaluate category, the algorithms use some technique such as a design of experiments (DOE) or a heuristic that will evaluate the initial small set of configurations and create new ones iteratively during execution. They repeatedly generate candidates and evaluate them, gathering historical information to better explore the search space. Works such as ParamILS~\cite{Hutter2009b}, SMAC~\cite{Hutter2011}, and GGA+~\cite{Ansotegui2015} are the state-of-the-art in the field of automatic parameter configuration. The final category refers to more recent mechanisms that will generate some high-quality configurations using a search method rather than by random-sampling or DOE methods. 

The AC approach considers the target algorithm of the optimization as a black-box problem since the relationship of the performance with the parameters is not available. Therefore, this kind of method needs to execute the algorithm to evaluate the performance of the chosen set of parameters. 
All the state-of-the-art methods try somehow to reduce the number of evaluations to find the best parameters quickly. All these methods can be applied successfully to find the best reservoir simulation numerical parameters, and many Bayesian Optimization and Genetic Algorithms are used in the O\&G for that purpose. However, despite not having access to the source code of the flow simulator, we propose that it is feasible to model its performance by the characterization of some features. Other works already explored this runtime prediction, saving considerable computational effort, as detailed in the next section.

\subsubsection{Runtime Prediction}

In~\cite{Buchaca2020} we proposed a mechanism to improve Spark\textregistered{} application performance by using SparkEventLog information from a single run of a workload to determine how its configurations should be tuned. We run it once to extract features, in a certain way similar to what a profiling tool does, characterizing the kind of workload in use. Then this information is used as the features of a regression model to predict the performance with different settings. All that was then needed was a search method, in this case a Bayesian Optimization (BO) being applied, but to the performance model that is almost ``instantaneous'' rather than the real workloads with the suggested parameters. This process was called Simulated Bayesian Optimization (SBO), as we query a performance model, instead of the real and expensive workload, to figure out the Spark configuration that will provide the lowest execution time. In~\cite{Buchaca2020} we also compared related works in six different categories: rule-based, cost modeling, simulation-based, experiment-driven, machine learning (or model-based), and adaptive tuning. We will not extend this discussion into this direction due to space constraints, but will focus below on the differences between the present work and~\cite{Buchaca2020}. 

These key ideas were the basis for TunaOil when working with reservoir simulations. Nevertheless, the direct application of that approach didn't work as well as with Spark and we discovered that the models do not generalize well. This makes sense because the shape of the reservoir and its characteristics play a very important role in the parameters of the reservoir model. Moreover, in this scenario, since we depend on physical measures, the amount of training data is much smaller than in other scenarios, such as an algorithm configuration. Therefore, we need a new methodology to tackle this problem, which is presented in Section~\ref{sec:proposal} by refitting the data within the loop of the HM process. It is also worth noting that the SBO relies on the classical Bayesian Optimization process with its acquisition function, which could be typically a Probability of Improvement (PI), an Expected Improvement (EI) or, an Upper Confidence Bound (UCB). This process of updating the acquisition points is computationally expensive and typically done in serial. As the search space of the reservoir simulator is much bigger than the one explored in the previous work, another significant change was to use another search method. As explained in Section~\ref{sec:query_oracle}, in this work we used a sampling technique to query the performance model instead of an optimization one as before.

The work \cite{Oladokun2020} is the closest work that we could find in the domain of reservoir simulation. They introduce a random forest regression that predicts the linear solver tolerance for each timestep of the simulation. This methodology is evaluated with two models: the SPE10 and a SAGD (thermal steam-assisted gravity drainage) model. Using an Oracle, they could reduce the linear iterations while keeping the nonlinear iterations near to the minimum. Despite not reporting the elapsed time of the experiments, the improved iteration counts suggest that it will speed up the simulation, and they report a ``good performance improvement''. A big difference from our work is that it has an ML system incorporated inside the simulator (which requires access to its source code), predicting the linear solver tolerance of the next timestep along with the simulation. Our approach was to define multiple numerical parameters that will be used constantly in the whole simulation. In that sense, the features to train the regression are very distinct from our work. They gather information from the previous timestep (e.g., the current timestep size or the number of linear iterations in the previous timestep) instead of the whole simulation (e.g., number of solver cuts or the timing on each kernel) as we propose. The work does not provide details about the training dataset or explain whether other regressions algorithms were explored and, if so, their performance.

\section{Problem Description}\label{sec:problem}

Our goal is to find a numerical parameter set for the reservoir simulator that reduces the execution time of the simulation of an ensemble of models that will be part of an ES-MDA without substantial degradation of the quality of the results. Let us define each of the parts in this optimization problem:

\begin{small}
\begin{itemize}
    \item $\mathcal{F}$ is the search space of the black-oil fluid model (the phase behavior, oil viscosity, etc.), $f \in \mathcal{F}$ is a particular instance. 
    \item $\mathcal{E}$ is the search space of the exploration strategy (wells, production schedule, etc.), $e \in \mathcal{E}$ is a particular instance.
    \item $\mathcal{G}$ is the search space of the geological model information (grid, fractures, etc.), $g \in \mathcal{G}$ is a particular instance. In the optimization workflow, we work with a single geological model ($g$), while in the HM workflow we use all the realizations ($g_{1},\dotsc,g_{\mathcal{N}_{r}}$).
    \item $\mathcal{S}$ is the search space of the numerical parameters of the reservoir simulator (the maximum number of Newton cycles allowed, the number of threads to use, etc.), $s \in \mathcal{S}$ is a particular instance of the search space. 
    \item $\mathcal{C}=\{q, b\}$ are the constraints on the optimization process, $q$ being a function that verifies the quality of the simulation output, and $b$ the total budget of the optimization process (a maximum number of iterations or a time limit).
    \item $t$ the elapsed time spent by reservoir simulator for a given input $(g,f,e,s)$. 
    \item $\hat{r}$ is a set containing all the metrics used for quality control (in our case, the results of the simulation such as the injection and production curves for oil, gas and water, plus the overall material balance for oil, gas, and water).
    \item $d_{obs}$ is the observed data to be compared with the simulated production data for performing the history matching.
    \item $d_{sim}$ is the simulator output production corresponding to the observed data $d_{obs}$. 
    \item $\mathcal{N}_{r}$ is the number of realizations in the geological ensemble.
    \item $\mathcal{N}_{a}$ is the number of assimilations performed by the ES-MDA.
\end{itemize}
\end{small}

\begin{figure*}[tb]
\centering
\includegraphics[width=1\textwidth]{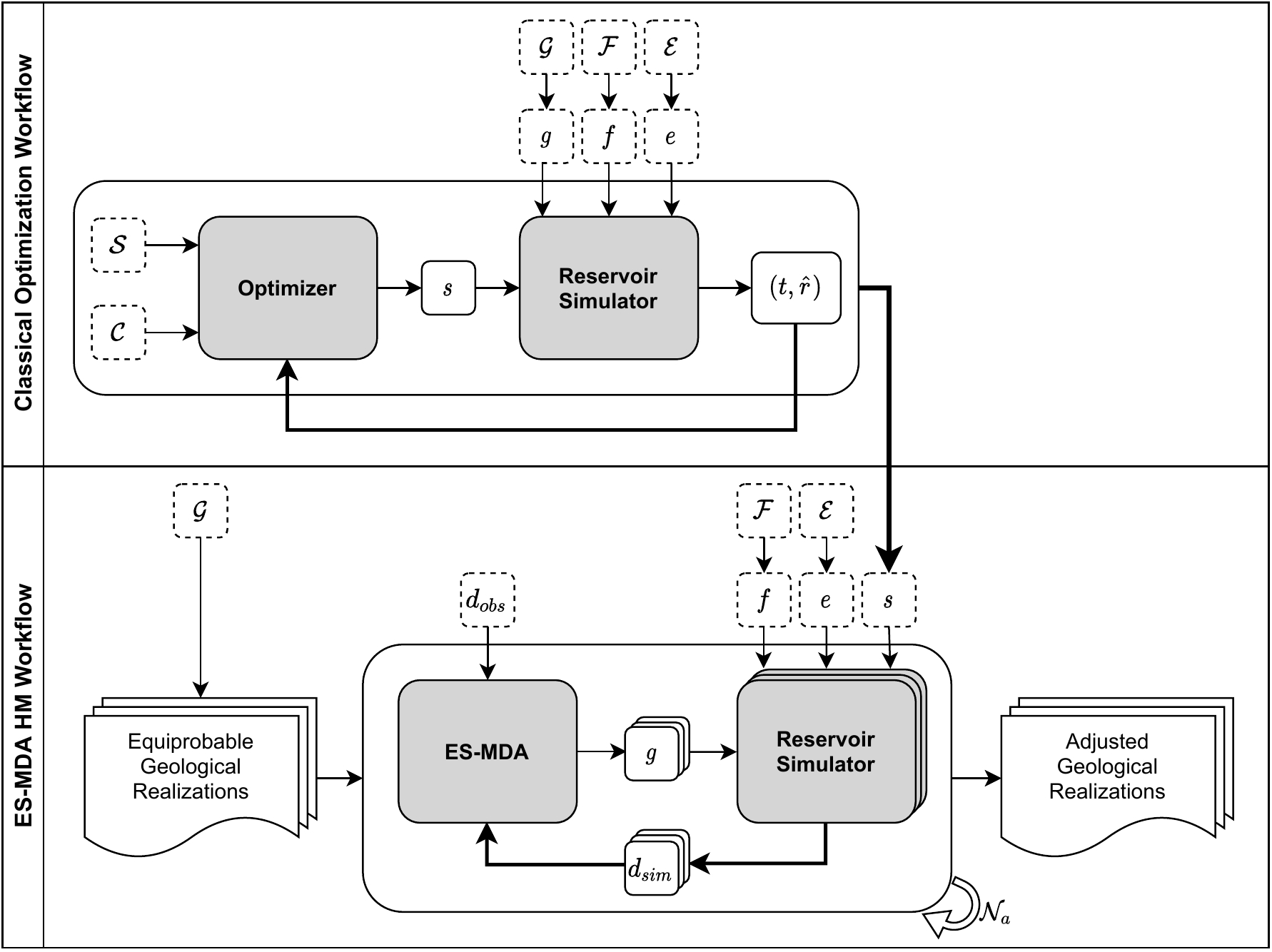}
\caption{The problem faced by the reservoir engineer as two independent workflows.}
\label{fig:problem}
\end{figure*}

The diagram in~\cref{fig:problem} illustrates the standard approach used to tackle the problem. It involves two completely independent workflows: the first, to optimize and find the best numerical parameters and the second, to run the HM workflow itself with the configuration previously found. The top of ~\cref{fig:problem} shows the first workflow, which consists of two applications represented by shaded boxes: the optimizer (e.g., a Bayesian Optimization or a Genetic Algorithm) and the numerical reservoir simulator. This classical optimization workflow uses a single geological model ($g$) to find the best numerical parameters ($s$). Subsequently, these fixed $s$ values are used for all of the realizations of the HM workflow using the ES-MDA. The dashed boxes represent fixed data that will not change during the execution of the workflows. The shadows represent multiple distinct data inputs/outputs or applications that are executed multiple times.

The diagram contains an optimization process that depends on the $\mathcal{S}$ search space and the constraints in $\mathcal{C}$. The overall optimization goal is to find an $s$ that reduces the simulation time, and that meets the quality constraints in $\mathcal{C}$. Note that $g$, $f$, and $e$ are fixed during the optimization process. A difficult decision is the definition of the particular geological model $g$ to use in the optimization. A common choice is to pick one of the models from the ensemble ($g_{1},\dots,g_{\mathcal{N}}$) at random or create a new model that represents the mean of all the models in the ensemble. In any case, the best parameter set found by the optimizer, which minimizes the simulation time without compromising the quality of the results, will only be truly the best for that particular geological model ($g$).

The second workflow is the HM itself. It consists of the execution of two independent applications also represented by two shaded boxes: the ES-MDA algorithm and the reservoir simulator, which are executed multiple times. The reservoir simulation will use the result of the numerical parameters ($s$) found previously for all the simulations to be performed, as the ES-MDA works with an ensemble of geological models. Therefore, it will not be a single particular geological model ($g$), but a set typically with hundreds of equiprobable geological realizations ($g_{1} \dots g_{\mathcal{N}_r}$). In each iteration of the ES-MDA, the whole set is simulated by the reservoir simulator. The result of all the $\mathcal{N}_r$ simulations are then used by the Kalman algorithm to change the ensemble of geological models ($g_{1},\dotsc,g_{\mathcal{N}_r}$) in such a way as to approximate the results ($d_{sim}$) to the observed data ($d_{obs}$). The updated ensemble ($g^{'}_{1}, \dotsc, g^{'}_{\mathcal{N}_r}$) will then be simulated again, and once more, the Kalman algorithm will perform the assimilation of the data (Equations~\ref{eq:EnKF}-\ref{eq:KalmanGain}). This specific Kalman algorithm is suffixed as the Multiple Data Assimilation (MDA) because it typically repeats this process four times. Four rounds of $\mathcal{N}_r$ simulations plus an extra final fifth round of simulations for the forecast ($\mathcal{N}_{a}=5$). The overall process might easily involve thousands of simulations \cite{Schulze-Riegert2019}.

\section{Proposed Solution}\label{sec:proposal}

\begin{figure*}[tb]
\centering
\includegraphics[width=1\linewidth]{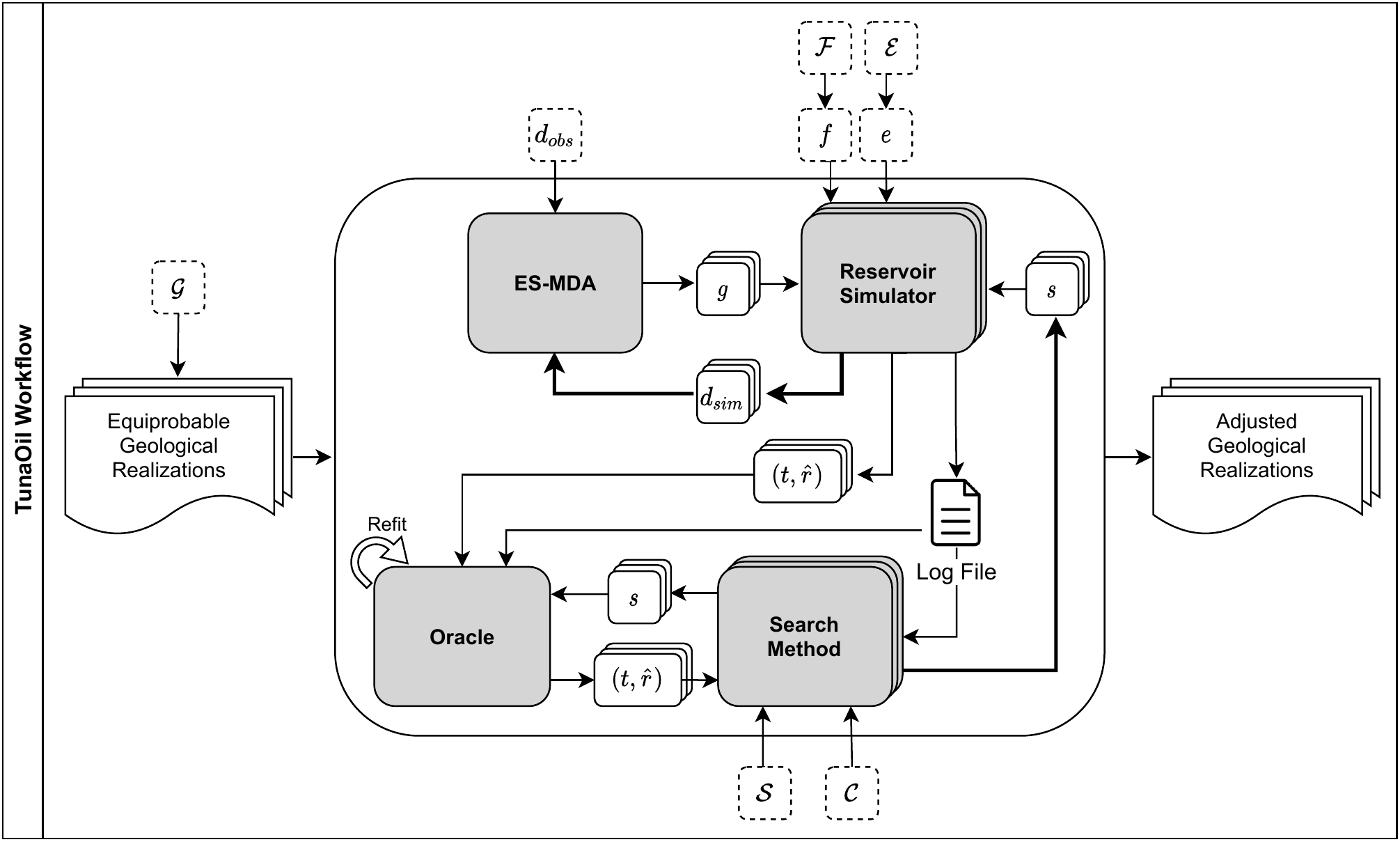}
\caption{Proposed solution workflow merging the history matching process with the search of the numerical parameters.}
\label{fig:proposed_solution}
\end{figure*}

To improve the overall process, we merged the two workflows into a single one using a performance model for quick selection of a numerical parameter set. As the ES-MDA is an iterative process, we can extract a feature vector from the output logfile from the previous reservoir simulations. This feature vector contains the information listed in~\cref{tab:nrs_features} and gathers fundamental knowledge about the underlying reservoir simulator execution. Therefore, our goal is to leverage this information to create an optimization-aware procedure that can speed up the whole process when compared to a classical optimizer that is not aware of such data (as in~\cref{fig:problem}). \cref{fig:proposed_solution} presents an overview of the proposed methodology.

The geology models are generated by the classical ES-MDA algorithm, while an Oracle provides the numerical parameters for the reservoir simulator. The first round of simulations uses the parameters provided by the engineer or the simulator defaults. In contrast, from the second iteration onward, the performance model uses the detailed log of the previous simulation as the feature vector for the predictions. The workflow in~\cref{fig:proposed_solution} is illustrated with separate processes simply to enable better understanding of the data flow. The query to the Oracle is done inside the ES-MDA loop, as better shown by the pseudo-algorithm \ref{alg:pseudocode}.

\begin{algorithm}[tb]
    \caption{Modified ES-MDA pseudo-code using TunaOil for tuning the numerical parameters of the ensembles.}\label{alg:pseudocode}
    \begin{small}
    \begin{algorithmic}[1]
    \Procedure{Coupled\_ES-MDA\_with\_TunaOil}{}
        \State Build the observed data array $d_{obs}$ of size $\mathcal{N}_d$
        \For {$i = 1$ to $\mathcal{N}_a$ } \Comment{until the number of assimilations defined}
            \If { $i > 1$ } \label{lst:line:tuna-begin}
                \State Refit Oracle with previous runs data \Comment{TunaOil Train}
                \State Get the best $s$ for each realization\Comment{TunaOil Predict}
            \EndIf\label{lst:line:tuna-end}
            \State Run the ensemble to collect $d_{sim}$
            \State Calculate $d_{uc}$ by disturbing $d_{obs}$
            \State Update the ensemble with the data from $d_{sim}$ as in Eq.\ref{eq:EnKF}
        \EndFor
        \State Run the final ensemble to obtain the forecast
     \EndProcedure
    \end{algorithmic}
    \end{small}
\end{algorithm}

The reservoir simulator input is essentially a text file (or a set of files) with the information about $g$, $f$, $e$, and $s$ defined by keywords in different sections. The set $(g,f,e,s)$ defines a unique reservoir simulation model. On the other hand, the output is composed of a text log file and a binary file. The log contains the details of the execution, warning and error messages, and some simulation outputs and statistics (such as the CPU profiling). One of the input keywords, in an I/O section, controls the level of detail output to the log. The binary file, in HDF5 format, has relevant information such as the metrics from $\hat{r}$ and all the updated maps of $g$ over time. \cref{fig:proposed_solution} is a simplified representation, and, in fact, we consume pieces of information from both output files.
The feature vector constructed, shown in \cref{tab:nrs_features}, comprises selected pieces of information from the grid, the exploration strategy, and the outputs from the reservoir simulation. 
The statistics in the description refer to the the minimum, maximum, mean, standard deviation, and discretized histograms for each curve or map. The CPU statistics in the features refers to the output in the log file of a special simulator flag (\texttt{{\small--cputime}}) that profiles, with almost no overhead, the time the simulator is spending on each main operation, such as the Jacobian, the Solver, Well Management, etc. This was shown to be one of the main features in the performance characterization.

\begin{figure*}[!htb]
   \centering
   \begin{tabular}{@{}lll@{}}
    \toprule
    {} & \thead{Reservoir Simulation\\Features} & \thead{Brief Description}  \\
    \midrule
 1 & \# Active Blocks       & The number of active cells in the grid         \\
 2 & \# Cuts                & Number of cuts in the timesteps                \\
 3 & \# Days Simulated      & The total number of days simulated in the run  \\ %
 4 & \# DOMS                & Number of domains for the decomposition       \\
 5 & \# Newton Cycles    & Total number of Newton Cycles performed       \\
 6 & \# Solver Failures  & Number of times linear solver failed to converge    \\ 
 7 & \# Timesteps        & Total number of timesteps performed           \\
 8 & \# Solver Iterations & The total number of linear solver iterations performed \\
 9 & \# Total Blocks        & The total number of cells in the grid (active + inactive) \\
10 & \# Wells               & Total number of wells (of any type)           \\
11 & $(ElapsedTime)/(\#TimeSteps)$  & Average Elapsed Time per Timestep             \\
12 & $\kappa$               & Statistics from the vertical and horizontal permeability map \\
13 & $\phi$                 & Statistics from the porosity map                \\
14 & $G_p$                  & Statistics on the gas production              \\
15 & $N_p$                  & Statistics on the cumulative oil production   \\
16 & $W_p$                  & Statistics on the water production            \\
17 & Average Implicitness   & The average implicitness of the matrix        \\
18 & CPU Statistics         & CPU time profiling output by the simulator    \\
19 & CPU Time               & The sum of time for all threads               \\
20 & Elapsed Time           & The wall time of the simulation               \\
21 & End of Simulation      & Status indicating if the simulation ends normally or with error \\
22 & MBE                    & The material balance error for each phase ($o, g, w$) \\
23 & Memory Usage Peak      & Total memory used in the simulation in MB     \\
24 & Simulation Time        & Number of years that the simulation is predicting (history + forecast) \\
25 & Simulator              & The reservoir simulator used with its version \\
    \bottomrule
    \end{tabular}
\captionof{table}{The feature vector extracted from past reservoir simulation executions.}
\label{tab:nrs_features}
\end{figure*}

The numerical parameter search space is given in~\cref{tab:search_space}. To select this subset of parameters, known as numerical keywords in the black-oil simulators, we use as a base the ones adopted in related works such as \cite{Avansi2019} and \cite{Rios2020}. The default configuration refers to the values used internally by the flow simulator when the user did not explicitly provide the given parameters. The parameter values are present in MODSI, a unit system based on the International System (abbreviated SI, from the French Système International d'unités), where the pressure-related variables use kg/cm\textsuperscript{2} instead of kPa as in the SI.

\begin{small}
\begin{table*}[htbp]
    \centering
    \footnotesize
    \begin{tabular}{@{}lllll@{}}
    \toprule
    \thead{Numerical Keyword} & \thead{Brief Description} & \thead{Search Space (MODSI)} & \thead{Space Encoding} & \thead{Default Configuration} \\
    \midrule
\begin{Code}
AIM
\end{Code}
& Adaptive implicit switching option & 
\begin{tabular}{@{}l@{}}
[OFF, STAB, \\
THRESH .25 .25 \\
STAB AND-THRESH] 
\end{tabular}
& Categorical & \scriptsize STAB AND-THRESH 0.25 0.25 \\
\begin{Code}
COMBINATIVE
\end{Code}
& Combinative 2-stage preconditioning & 
[OFF, ON, AMG, ILU] & Categorical & OFF \\ 
\begin{Code}
DTMAX
\end{Code}
& Maximum time-step size allowed & 
[5, 365] & Real & 365 \\
\begin{Code}
DTMIN
\end{Code}
& Minimum time-step size allowed & 
[1E-6, 0.001] & Real & 0.001 \\
\begin{Code}
ITERMAX 
\end{Code}
& Maximum linear solver iterations & 
[5, 200] & Integer & 10 \\ 
\begin{Code}
MAXCHANGE PRESS
\end{Code}
& Maximum variation in pressure per time-step & 
[15, 300] & Integer & 60 kg/cm2 \\
\begin{Code}
MAXCHANGE SATUR
\end{Code}
& Maximum variation in saturation per time-step & 
[0.1, 0.9] & Real &  0.1 \\ 
\begin{Code}
NCUTS 
\end{Code}
& Maximum number of time-step cuts & 
[None, 10, 100] & Integer & None (Unlimited) \\
\begin{Code}
NEWTONCYC 
\end{Code}
& Maximum number of Newtonian cycles & 
[5, 40] & Integer & 10 \\
\begin{Code}
NORM PRESS
\end{Code}
& Normal variation in pressure per time-step &
[10, 70] & Integer & 30.0 kg/cm2 \\
\begin{Code}
NORM SATUR
\end{Code}
& Normal variation in saturation per time-step &
[0.1, 0.3] & Real & 0.1 \\
\begin{Code}
NORTH
\end{Code}
& Maximum number of orthogonalizations & 
[15, 200] & Integer & 30 \\
\begin{Code}
PIVOT
\end{Code}
& Diagonal sub-matrix inversion pivot stabilization & 
[OFF, ON] & Categorical & OFF \\
\begin{Code}
PRECC
\end{Code}
& Convergence tolerance for linear solver & 
[$10^{-5}$, $10^{-3}$] & Real (Log-Uniform) & $10^{-4}$ \\
\begin{Code}
SOLVER
\end{Code}
& Linear equation solver to be used & 
[AIMSOL, PARASOL] & Categorical & AIMSOL \\
\begin{Code}
SORDER
\end{Code}
& Solver equation ordering in the ILU factorization & 
\begin{tabular}{@{}l@{}}
[NATURAL, RCM, \\
REDBLACK, RCMRB]
\end{tabular}
& Categorical & REDBLACK \\
    \bottomrule
    \end{tabular}
    \caption{The search space of the different parameters for the simulator used in this work with the respective default values.} 
    \label{tab:search_space}
\end{table*}
\end{small}

\subsection{Query of the Oracle}\label{sec:query_oracle}

The performance model works by querying the Oracle with all the features extracted from the reservoir simulation logs (\cref{tab:nrs_features}), plus a candidate numerical configuration ($s$) belonging to the search space (\cref{tab:search_space}). As a result, it returns a prediction of the elapsed time and quality for that particular configuration ($s$) on the given reservoir model, characterized by the feature vector. We typically use an optimization method to search into the parameter space to find a suitable parameter set. 

In a previous work~\cite{Buchaca2020} we followed that approach by using a Bayesian Optimization (BO) method to query the performance model in what we called a Simulated Bayesian Optimization (SBO). The SBO converged quickly on a good solution that minimizes the execution time of our workload - a Spark job. The Spark problem already had an excellent converged solution with 20 iterations with the classical Bayesian Optimization (BO-20). For the SBO with Spark, we did 100 BO iterations querying  the Oracle, which took only 2 seconds. Our first query idea was to repeat the SBO approach from the work~\cite{Buchaca2020} with the reservoir simulation workloads. However, the first issue we faced was that the BO/SBO required many more iterations to find a good parameter configuration with the reservoir simulation workloads - in the order of two hundred to five hundred iterations. This more significant number of iterations could be a consequence of the respectively bigger parameter space shown in~\cref{tab:search_space}. The larger number of iterations is a problem for BO, as it requires the execution of the real workload, but should not be a problem for the SBO when it is just querying the Oracle. However, the SBO still needs to update (fit) the internal probability model (the surrogate). This fit process is quick (less than a second) for the first iterations. However, when the convergence is challenging due to the vast search space, it starts to be exponentially slower (minutes) for hundreds of iterations, which led us to think about some alternatives. 
One naive idea was to generate random samples of the parameter configuration ($s$) to overcome the slowness of the Bayesian Optimization update of its internal acquisition function. The performance model response is almost instantaneous (less than one second), and we can query multiple samples at once without performance degradation. Therefore, if we generate thousands of random samples, we have a higher chance of finding suitable configurations, although it will not be guaranteed that the best configuration encountered is the absolute minimum. All optimization methods face the same issue, as we need to set a budget that can be the number of iterations or a wall time constraint for the whole process. We are interested in finding a ``good enough'' set, better than the defaults and effortless for the engineer. We can use a clever sampling technique rather than random sampling to improve this approach, and we chose the Latin Hypercube Sampling method for this purpose.

Latin Hypercube Sampling (LHS) is a statistical method that aims to spread the sample points more evenly across all the space of possible values. It generates quasi-random samples of the parameter values from a multidimensional search space. The samples are quasi-random because the method knows beforehand how many random points will be sampled and can spread these points, ensuring the exploration of each parameter dimension. For a better understanding of LHS, let us consider a two-dimensional space. Latin Square Sampling will divide the square space, formed by the two variables, into a $n x n$ chess mesh, and choose $n$ points spread throughout each sub-square, but in such a way that each line or column has one, and only one, sample. LHS generalizes this method for multiple dimensions. Some optimization techniques, such as BO, require one sample result to be evaluated to suggest another sample sequentially. In contrast, LHS is highly parallelized and can generate all the samples at once. In our case, the query of the Oracle with 10,000 parameters $s$ was generated by an LHS, and the other required data was extracted from the previous ES-MDA iteration.

\subsection{Prediction Results}

To find the best numerical parameters, the Oracle needs to predict the simulation time and quality correctly, or at least the impact of the parameters on these two metrics. 
Although the ML predictions are often distant from the reality, the expected behavior  on how one parameter impacts the time or quality is being correctly predicted. 
In other words, even when predicting a wrong elapsed time in comparison with the real values obtained by a reservoir simulation, the regressor was capable of indicating that a given parameter will slow down or speed up the simulation. The same applies to the MBE prediction when comparing two predictions with different parameters for the same reservoir model. In any case, we need to deal with the two prediction results, and as a result, we decided to  weight the elapsed time accordingly to the quality of the model to choose which of the 10,000 LHS results will be selected. The lower the MBE, the better. Therefore, for up to 5\% we use the raw value of the elapsed time. For higher errors, up to 10\% in the prediction, we apply a penalty. And more than that is unacceptable from the engineering point of view. 
As a result, we adopted a weighted elapsed time to rank the results and select the best for each reservoir model, as defined by:

\begin{equation}\label{eq:WET}
WET = 
\begin{cases}
ET        & \because \max(\overline{MBE}) \leqslant 0.05 \\ 
ET * 2    & \because 0.05 < \max(\overline{MBE}) \leqslant 0.10 \\ 
ET * 1000 & \because \max(\overline{MBE}) > 0.10 
\end{cases}
\end{equation}
where:
\begin{itemize}
  \item $WET$ stands for Weighted Elapsed Time
  \item $ET$ stands for Elapsed Time
  \item $\overline{MBE}$ stands for the average of the Material Balance Error among the phases: $(MBE_o+MBE_w+MBE_g)/3$
\end{itemize}

\subsection{Proposal Considerations}

Reservoir models can perform very differently due to factors that are very difficult to characterize, such as the degree of heterogeneity and non-linearities resulting from the characteristics of the problem, such as flow rates, mass transfer between phases, etc. Using the first iteration of the ES-MDA to retrain the Oracle gives more control over the flow simulation to give correct predictions, as only the maps of the properties such as the permeability and porosity will change between the HM iterations. These simulations tend to be  very similar, unlike a distinct reservoir model never seen before by the Oracle. Following this approach, we also guarantee that we will not have parameters that lead to the simulation not converging in the first iteration due to some model particularity. Therefore , the risk of breaking the ES-MDA is significantly reduced by the refit using the first iteration data. Another consideration relates to the simulator supporting multiple systems of units. Many inputs and outputs of the simulator may be associated with some fundamental measurement units (e.g., the temperature, which may be measured in Celsius or Fahrenheit). Special care was taken to keep them consistent for the performance model to be trained correctly.

\section{Workloads}\label{sec:workloads}

There are two groups of workloads used in this work: one for training the performance model and the other for in-depth validation of the Oracle. 

\subsection{Reservoir Models for Training}\label{sec:workloads:initial-dataset}

To build the initial dataset used to train the performance model, around 25,000 black-oil models were generated from reservoirs models,  fourteen being from real fields and two from the literature (the Brugge field\cite{Peters2010} and Model 2 of the Tenth SPE Comparative Solution Project\cite{Christie2001}, also known as SPE10 Model 2). Some of these reservoir models have multiple realizations of a 3D geological model, while others have a single model. For instance, Brugge consists of 104 scaled-up realizations, while SPE10 is just one model. These reservoir models have different sizes, in terms of cells, different fluid models (despite all being black-oil), different numbers of wells, etc. Therefore, they also have very different simulation times. Table~\ref{tab:workloads_oracle} summarizes this dataset, the cases Brugge and SPE10 being the workloads \#3 and \#16, respectively.

To give a sense of the complexity of each model, we added two metrics to Table~\ref{tab:workloads_oracle}: the Nonlinear Iterations per Time Step (NI/TS) and the Linear Iterations per Nonlinear Iteration (LI/NI). The lower these numbers, the easier it is for the simulator to find the solution for the problem it is trying to solve. Those metrics are not directly reflected in the elapsed time, which depends on the beginning and end simulation time for each model as well as the number of timesteps each will have.

\begin{small}
\begin{table*}[tb]
  \centering
  \begin{tabular}{rrrrrrrrrrrrrrrr}
  \toprule
    \multirow{2}{*}{\textbf{ID}} & \phantom{} & \multicolumn{2}{c}{\textbf{Cells}}  & \phantom{} & \multirow{2}{*}{\textbf{Wells}}  & \phantom{} & \multicolumn{3}{c}{\textbf{Defaults}} & \phantom{} & \multicolumn{3}{c}{\textbf{Engineer}} & \phantom{} & \multirow{2}{*}{\textbf{Configs}} \\
    \cmidrule{3-4} \cmidrule{8-10} \cmidrule{12-14} 
  & & \textbf{Active}  & \textbf{Total} & &  & & \textbf{NI/TS} & \textbf{LI/NI} & \textbf{ET} & & \textbf{NI/TS} & \textbf{LI/NI} &  \textbf{ET} & &  \\
  \midrule
   1 & &    28,971 &     85,905 & &  25 & & 1.43 & 29.19 &     1 & & 2.26 & 33.31 &    2 & & 1,164 \\ %
   2 & &    40,354 &    788,242 & &  90 & & 1.68 & 26.19 &     8 & & 1.65 & 36.36 &    9 & & 1,534 \\ %
   3 & &    43,217 &     60,048 & &  30 & & 1.05 & 39.37 &  $<$1 & & 1.05 & 39.87 & $<$1 & & 1,715 \\ %
   4 & &    55,863 &    278,244 & &  20 & & 7.26 & 36.26 &    26 & & 7.51 & 39.30 &   26 & & 1,787 \\ %
   5 & &   128,686 &  1,301,643 & &  36 & & 4.09 & 21.99 &    83 & & 1.64 & 30.22 &   74 & & 1,239 \\ %
   6 & &   261,280 &    659,520 & &  10 & & 4.34 & 26.10 &    36 & & 3.76 & 33.48 &   27 & & 1,798 \\ %
   7 & &   284,046 &  2,113,848 & &  46 & & 2.56 & 20.26 &   153 & & 3.55 & 14.48 &   84 & & 1,168 \\ %
   8 & &   334,365 &    630,400 & &  12 & & 2.76 & 20.87 &    13 & & 2.30 & 40.71 &   16 & & 1,793 \\ %
   9 & &   415,929 & 10,531,206 & & 117 & & 3.37 & 28.89 &   164 & & 3.09 & 37.56 &   95 & & 1,209 \\ %
  10 & &   540,747 &  4,599,672 & &  32 & & 5.67 & 28.22 &   104 & & 7.27 & 36.34 &   73 & &   994 \\ %
  11 & &   585,099 &  1,300,320 & &  26 & & 4.65 & 16.29 &    73 & & 3.42 & 33.41 &   60 & & 1,154 \\ %
  12 & &   607,113 &  1,803,420 & &  26 & & 5.02 & 22.05 &    71 & & 4.38 & 37.47 &   59 & & 1,638 \\ %
  13 & &   733,534 &  4,752,360 & &  26 & & 5.48 & 11.52 &   244 & & 5.37 & 24.56 &  233 & & 1,084 \\ %
  14 & &   765,620 &  1,491,952 & &  89 & & 2.79 & 8.82  &   258 & & 2.66 & 4.78  &  312 & & 1,721 \\ %
  15 & &   880,796 &  4,752,360 & &  26 & & 5.02 & 15.70 &   225 & & 5.24 & 26.78 &  212 & & 1,163 \\ %
  16 & & 1,094,421 &  1,122,000 & &   5 & & 6.92 & 39.94 &   131 & & 7.53 & 40.00 &  125 & & 1,435 \\ %
  \midrule
  \multicolumn{15}{l}{TOTAL} & 22,540 \\
  \bottomrule
  \end{tabular}
  \caption{Reservoir models used to train the Oracle. NI/TS stands for Nonlinear Iterations per Time Steps, LI/NI stands for Linear Iterations per Nonlinear Iterations, and ET stands for the mean Elapsed Time in minutes for 48 cores. The number of configurations consider only the valid ones (that performed a full simulation). All configurations were executed more than once and the values present are the mean.}
  \label{tab:workloads_oracle}
\end{table*}
\end{small}

As the main objective is the performance prediction, one important consideration was to ensure that the computational environments were the same between runs when measuring all time executions. More details about the hardware architecture are given in Section~\ref{sec:experiment:computational_environment}. 
Furthermore, each unique configuration (a geological model with a given numerical parameter set) was executed multiple times (workloads \#13, \#14 and \#15 were run only twice as they were the slowest, but all others three times). 
Nevertheless, the total number of simulations in Table~\ref{tab:workloads_oracle} refers to unique combinations of the numerical parameters with the geological, fluid, and exploitation models (but does not consider the repetitions).
The ``Engineer'' time in ~\cref{tab:workloads_oracle} refers to the elapsed time using the numerical parameters defined by a reservoir engineer and in use in their respective production environments. The engineer generally uses a Classical Optimization Workflow, as illustrated in \cref{fig:problem}. However, we don't have the proper tracking of which optimization techniques were used to elect those setups as the ``best'' for those models. The optimization technique could range from a Particle Swarm technique to brute force optimization or even manual tuning based on engineer experience. An issue of using these parameters as a baseline is that the engineer could have used different ``objective functions'' rather than the elapsed time, for instance, to lower the MBE. Likewise, the ``Default'' time uses the default values provided by the reservoir simulation (equivalent to not providing any numerical configuration at all). 
The number of simulations for a given workload represents a unique combination of the numerical section parameters. Each unique configuration was simulated more than once and the elapsed times used on the train/test of the Oracle are the mean of these runs. 

In addition to the execution of the two baselines (engineer and default), we were required to execute each reservoir model with multiple numerical parameter combinations to have a diverse dataset for the training. We employed three different techniques with all the models to try to capture the influence of the parameters. The first was a ``One-at-a-Time'' (OaT) Sensitivity Analysis that varies one parameter among the defined values in the space while keeping all others fixed. The second was Bayesian Optimization. And the last was Latin Hypercube Sampling. All of them operated within the search space given in \cref{tab:search_space}. The number of valid configurations shown in~\ref{tab:workloads_oracle} refers only to the configurations that were used for the training. Many simulations were discarded from the dataset due to errors or timeouts, as explained in \cref{sec:experiments:performance-model-evaluation}. We defined a timeout as $2 \times max(default, engineer)$ for all those executions, as some poor parameters configurations could lead to very long simulations.

\subsection{Reservoir Models for Validation}\label{sec:workloads:validation-models}

To evaluate this work, the three reservoir models listed in Table~\ref{tab:workloads_validation} were considered. These models were not part of the initial dataset described in \cref{sec:workloads:initial-dataset}, so our performance model did not know them beforehand. They were chosen because they are public model ensembles used by many academic studies and competitions, facilitating the reproducibility and future comparisons of our work. The ES-MDA workflow runs each realization five times, changing the permeability and porosity maps. This section will describe all of the models for the experiments in further detail.

\begin{table}[h!tb]
 \centering
  \begin{tabular}{lcc}
  \toprule
  \textbf{Workload   } & \textbf{\# Realizations} & \textbf{\# Simulations} \\
  \midrule
  \texttt{OLYMPUS}       & 50                       & 250                     \\
  \texttt{UNISIM-I}      & 48                       & 240                     \\
   \texttt{UNISIM-II}  & 500    & 2,500          \\
  \bottomrule
  \end{tabular}
  \caption{Reservoir models for cherry-pick validation.}
  \label{tab:workloads_validation}
\end{table}

\textbf{OLYMPUS} is a synthetic reservoir model inspired by a real field in the North Sea and developed by TNO in 2017 for a benchmark study on field development optimization \cite{Tno2018,Fonseca2020}. This challenge encompasses three different tasks: a well control optimization, a field development optimization, and a complete closed-loop exercise (a joint field development and well control optimization). The model created by TNO has enough complexity, as a real field, to cover all these problems with an ensemble of 50 geological realizations for rock-fluid properties (porosity, horizontal and vertical permeability, net-to-gross, and initial water saturation). The field is 9 by 3km wide with thickness of 50m, modeled in 16 layers as a grid of 341k cells (192,750 active), one of the layers being a shale barrier forming most of the inactive cells in the model. \cref{fig:workload_olympus} shows the permeability in layer 1 from three different realizations with the distribution of the wells. A full description of the challenge problems and the model details can be found in \cite{Tno2018}. 

While a single simulation of OLYMPUS can execute in less than 5 min (with the environment described in \cref{sec:experiment:computational_environment}), it is interesting to note that to solve some of the problems posed by the challenge could require a computational effort of more than 1,000 simulations, as reported in \cite{Schulze-Riegert2019}. Summing all the solution strategies that these authors applied to the challenge TNO proposed, it required them 18,900 simulations of OLYMPUS models. This massive number of simulations shows that some workflows may have to run the ``same'' reservoir model so many times that any slight improvement in the execution times of those models more than compensates for the training cost required by our approach, even for relatively quick simulations. OLYMPUS presents a NI/TS of 1.43 and a LI/NI of 120.90.

\begin{figure}[tb]
\centering
\includegraphics[scale=0.20,trim=140 185 10 180,clip]{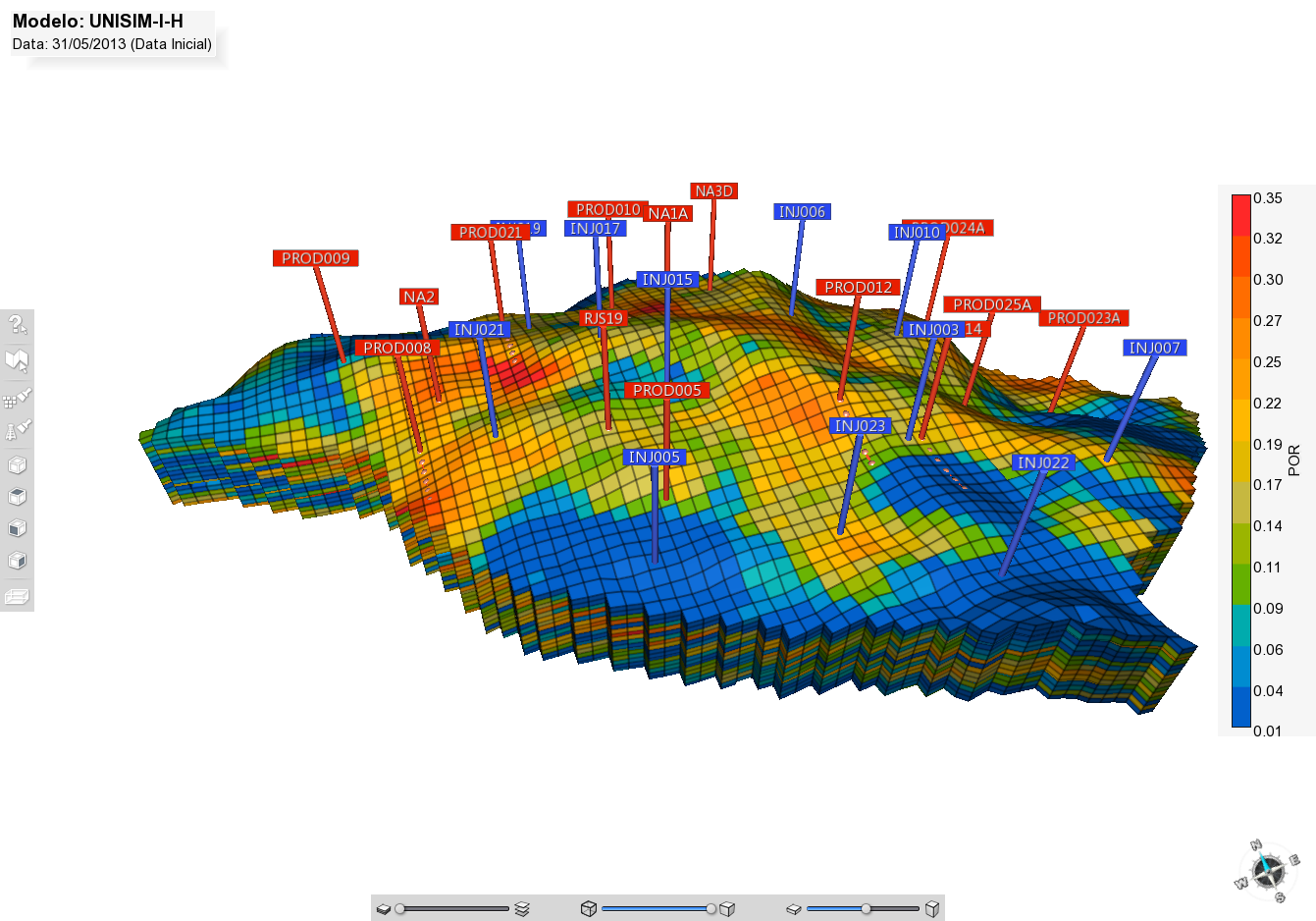}
\caption{Porosity 3D map of one realization of UNISIM-I upscaled model.  The injectors (blue) and producers (red) wells are also displayed.}\label{fig:workload_unisim-i-r_upscaled}
\end{figure}

\textbf{UNISIM} is a set of synthetic models developed by UNICAMP and built based on various off-shore fields on the Brazilian coast. The UNISIM benchmarks used in this work were the \textbf{UNISIM-I} and \textbf{UNISIM-II}. Despite similar names, they are very different models. \textbf{UNISIM-I}, developed by \cite{Avansi2015}, is based on the Namorado Field. The base reference model, known as UNISIM-I-R, is discretized in a corner-point grid of $326{\times}234{\times}157$ cells, which lead to a total of almost 12M cells (3,408,633 active), with a cell resolution of $25{\times}25{\times}1m$. The UNICAMP group also created the UNISIM-I-D model for a project development scenario with uncertainties, providing equiprobable geological realizations. This uncertainty set has been upscaled to $81{\times}58{\times}20$, reducing the total size to 93k cells (36,739 active), with a cell resolution of $100{\times}100\times8m$. \cref{fig:workload_unisim-i-r_upscaled} shows the 3D structure of the upscaled reservoir model with the wells. UNISIM-I-R presents an NI/TS of 1.83 and an LI/NI of 26.26 for the upscaled version, respectively and 3.47 and 23.35 for the fine one. These values are the average among all realizations.

\begin{figure}[tb]
\centering
\includegraphics[scale=0.21,trim=160 150 10 170,clip]{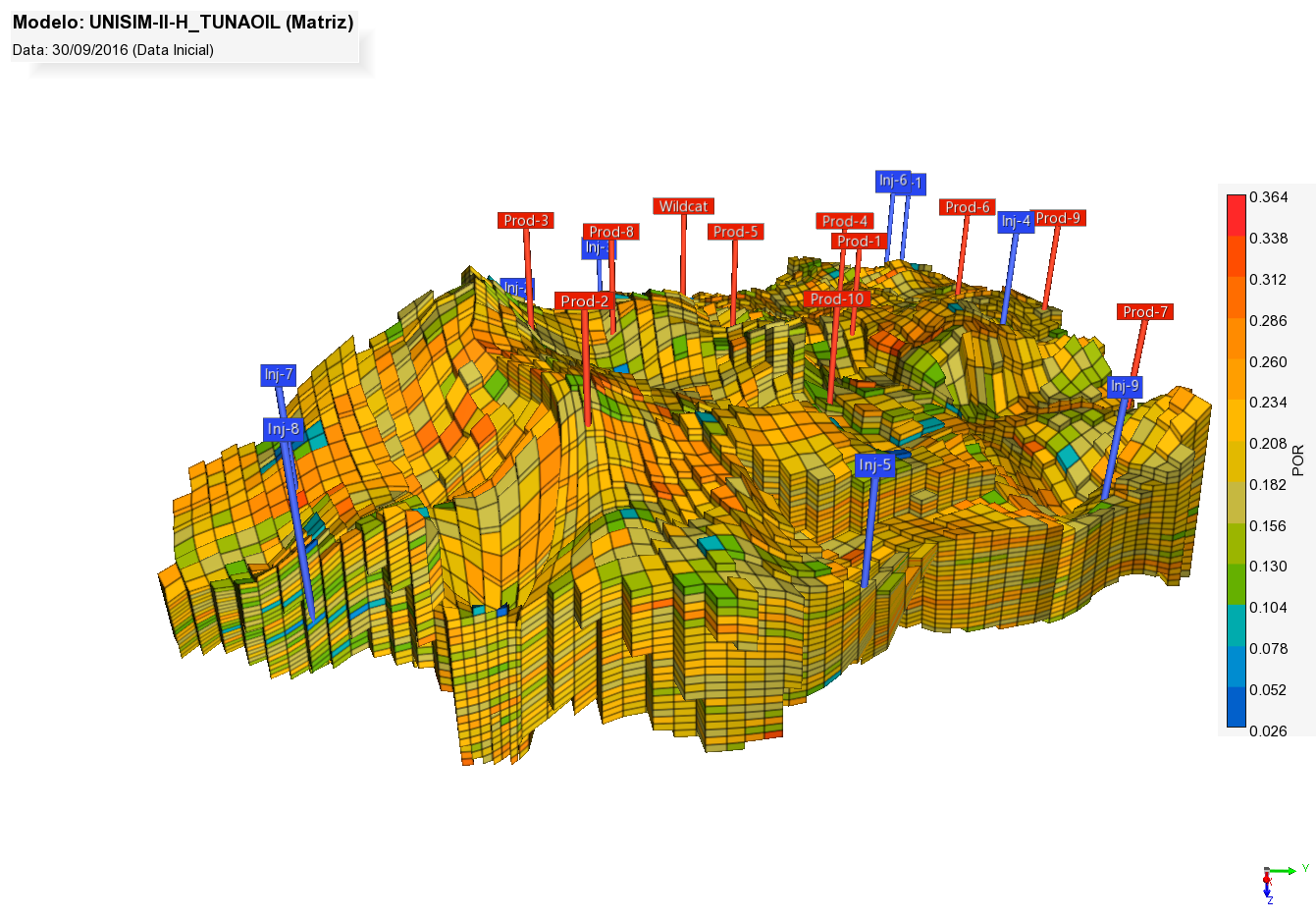}
\caption{3D porosity map of a UNISIM-II-H realization with its wells (injectors in blue and producers in red).}\label{fig:workload_unisim-ii-h_upscaled}
\end{figure}

The \textbf{UNISIM-II} benchmark model, developed by \cite{Correia2015}, represents a carbonate oil field combining the pre-salt and Ghawar fields information. The authors provide different models and the chosen one for our validation was UNISIM-II-H\cite{Maschio2018}, which is specifically intended for history matching and uncertainties reduction. It has a grid with 190,440 total cells with average dimensions of $100{\times}100{\times}8$m. A double-porosity, double-permeability representation is used to model the fracture behavior. Each cell has values related to 5 properties: matrix permeability, fracture permeability, matrix porosity, fracture porosity, and net-to-gross. The case includes 500 geostatistical realizations (petrophysical property maps), representing possible spatial distributions of the five properties, taking into account the probability distribution of each one. Depending on the realization, the active cells ranged from 59,405 to 81,551 active blocks. \cref{fig:workload_unisim-ii-h_upscaled} shows the 3D mesh of one realization of the upscaled reservoir model with its 20 wells, being 9 horizontal water injectors and 11 vertical producers. The model presents high vertical heterogeneity in permeability with high permeability thin zones, in a fractured oil reservoir with 16 faults. They provide 3,257 days of historical production data and  10,957 simulation days in total. The model presents an average NI/TS and LI/NI of 1.39 and 33.06, respectively, considering all realizations.

TNO provided the original OLYMPUS simulation input files in the format of the black oil simulator Eclipse 100\footnote{https://www.software.slb.com/products/eclipse} from Schlumberger and later a Petrobras engineer converted them to the IMEX\footnote{https://www.cmgl.ca/imex} simulator from CMG. The UNISIM models were already available on their web page\footnote{https://www.unisim.cepetro.unicamp.br/} in CMG format, not requiring any conversion. While both black-oil simulators are conceptually similar, they have internal differences in their algorithms that require different numerical tuning to reach the same results, besides having different default values. In the conversion of the OLYMPUS model to CMG format, the engineer chose to prioritize the match of the curve results and eliminate any material balance error. This choice will have some performance consequences, as will be shown in the \cref{sec:experiments:performance-model-validation}.

Despite executing the black oil simulator IMEX for all the experiments, OLYMPUS uses a simpler fluid model by using the keyword \texttt{*MODEL *OILWATER} which restricts the internal calculation to a two-phase, oil and water model, with no modeling of free gas or variation in solution gas. In terms of numerical complexity, this turns it into the simplest of the three reservoir models as the solver only receives two equations per grid block. Both UNISIM workloads use the keyword \texttt{*MODEL *BLACKOIL}, which leads to a more complex fluid calculation because of the internal modeling of oil, water and gas phases. UNISIM-I extrapolation strategy, as used in this work, keeps the pressure of the reservoir above the bubble point due to the action of water injector wells, which avoids the generation of free gas during the simulations. UNISIM-II presents the most complex physics calculation of the three, as besides the presence of free gas, it uses a double-porosity, double-permeability model with fracture, which leads to the calculation of the inter-block, matrix-to-matrix flows.

\section{Experiments}\label{sec:experiments}

To assess the improvement from our TunaOil approach, we raise a series of Research Questions (RQ) in this section that lead to a series of experiments to answer them.

\begin{itemize}
\item \textbf{RQ1}: Does the proposed feature vector characterize the performance of a reservoir simulation in accordance with different numerical parameters? 
\item \textbf{RQ2}: Can an Oracle trained with these features predict the performance and quality of a simulation with an unseen reservoir model? 
\item \textbf{RQ3}: How much performance improvement, with respect to the default simulator parameters, can TunaOil achieve? 
\item \textbf{RQ4}: How much performance improvement, with respect to manual engineer optimized parameters, can TunaOil achieve?
\end{itemize}

To answer \textbf{RQ1} and \textbf{RQ2}, we detail in~\cref{sec:experiments:performance-model-evaluation} the results of the train-test of the performance model. Furthermore, we demonstrate the methodology used by splitting the data into partitions to ensure that we never evaluate the performance model on samples of reservoir data seen during the training. For \textbf{RQ3} and \textbf{RQ4}, we consider the models with the default and engineer numerical parameters, respectively, as the baselines to measure the TunaOil improvement in an ES-MDA evaluation. \cref{sec:experiments:performance-model-validation} shows how these compare with the reservoir models OLYMPUS, UNISIM-I and UNISIM-II, detailed previously in~\cref{sec:workloads:validation-models}.

\subsection{Performance Model Evaluation}\label{sec:experiments:performance-model-evaluation}

This section details how the data was prepared, how the model was constructed, and the results obtained:

\textbf{Data Processing}: The first step was to clean the data by performing some pre-processing on the raw data. We removed all the simulations that do not present a ``Normal Termination'' from the training dataset. Initially, we tried to set a large elapsed time for any "Abnormal Termination" or "User Interrupt" cases, but the prediction errors  were higher with that approach. We also removed simulation cases with a single timestep (for some reason, ``Normal Terminations'' with this situation occured a few times). 

We also applied some transformations to the features. We used one-hot encoding\footnote{\MYhref{https://scikit-learn.org/stable/modules/generated/sklearn.preprocessing.OneHotEncoder.html}{scikit-learn OneHotEncoder reference}} to transform all the categorical data into their numerical counterparts. This preprocessing technique transforms each possible category value into a new column and assigns a zero (false) or a one (true) depending on whether the data has the given value. For the rest of the variables, we applied Rescaling\footnote{\MYhref{https://scikit-learn.org/stable/modules/generated/sklearn.preprocessing.StandardScaler.html}{scikit-learn Rescaling reference}} or Standardization\footnote{\MYhref{https://scikit-learn.org/stable/modules/preprocessing.html}{scikit-learn Standardization reference}}. The former transforms the data into the $[0,1]$ range, while the latter normalizes the values by removing the mean and applying unit variance. We opted to construct a pipeline\footnote{\MYhref{https://scikit-learn.org/stable/modules/compose.html}{scikit-learn Pipeline reference}} including these two transformations. Accordingly, each regressor was evaluated with both and the decision on which transformation to use in the final trained model is delegated to the cross-validation. As we wanted to explore different regressors, this seemed a better decision as some learning algorithms (such as Support Vector Machines) assume the same behavior is used for all features. If Standardization had not been performed, we could have one feature dominating the objective function if its variance is orders of magnitude larger than other features \cite{scikit-learn-preprocessing}. However, including the transformations of the non-categorical data in the pipeline increases the training time as the grid-search evaluates each regressor twice. Nevertheless, it will lead to better results with the best transformation being applied to each learning algorithm.

We also included a Feature Selection stage in the pipeline, using the highest $k$ scores as the selection strategy. We added three options to the grid search for $k$: 100\%, 90\%, and 80\% of the features. Accordingly, we could evaluate the results using all the feature vectors defined initially, as well as with reduced subsets. The reduction can improve the computational cost of the Oracle, but the more significant benefit is the possibility of improving the results of the predictions. As we included it in the pipeline, it simplifies future queries of the performance model, as we could keep using all the original features.

\textbf{Model Evaluation}: We used an approach known as Leave One Group Out Cross-Validation (LOGO-CV)\footnote{\MYhref{https://scikit-learn.org/stable/modules/cross_validation.html}{scikit-learn LOGO-CV reference}}, which works with a provided information group and splits the data accordingly. This group information allows us to leave out all the simulations of one given reservoir for a better validation. As there will be less variation in the elapsed times between simulations using the same base reservoir model, this methodology avoids data leakage. Additionally, LOGO-CV provides better results when compared to standard Cross-Validation (CV), as we are evaluating never-seen models. Therefore, the resulting prediction model will generalize better. In our case, LOGO-CV splits the dataset into 16 different groups, using the ID shown in~\cref{tab:workloads_oracle} to leave out one of each set of reservoir models in each split. In the end, the average of the errors from all the validation partitions was computed and reported. 

\textbf{Model Selection:} We used a grid search to train ten estimators using different parameter values, as shown in~\cref{tab:hyperparameters}. Each estimator was trained 16 times with LOGO-CV for each of its hyper-parameters. This process was very costly due to the high train time of some regressors, especially the Support Vector Regression (SVR), the Multi-Layer Perceptron (MLP), and the Gradient Boosting (GB) regressors. The overall grid search took almost three days using all the cores of a node of the BSCDC cluster described in~\cref{sec:experiment:computational_environment}.

\begin{table}[!htbp]
\centering
\begin{small}
\begin{tabular}{ll@{}ll@{}}
    \toprule
    \thead{Model} & \thead{Hyper-parameter Search Space}               \\
    \midrule
    Decision Tree & criterion: [``mse'', ``friedman\_mse'', ``mae'']   \\
    Elastic Net   & alpha: $[1e^{-10},1e^{-9},\dotsc,30.0]$            \\
    Extra Tree    & criterion: [``mse'', ``friedman\_mse'', ``mae'']   \\
    GB            & learning\_rate = [0.001, 0.01, 0.1, 0.5]           \\
                  & max\_depth =[3, 10, 20, None]                      \\
    KNN           & n\_neighbors: [1, 2, 3, 4, 5, 6]                   \\
                  & p: [1, 2]                                          \\
    Lasso         & alpha: $[1e^{-10},1e^{-9},\dotsc,30.0]$            \\
    MLP           & activation: [``tanh'',``relu'']                    \\
                  & hidden\_layer\_sizes: [[100], [200], [250], [300], \\
                  & [200,200], [250,250,250], [250,250,250,250]]       \\
                  & learning\_rate: [``adaptative'']                   \\
                  & max\_iter: [250]                                   \\
                  & solver: [``lbfgs'', ``sgd'']                       \\
    Random Forest & max\_depth: [10, 20, None]                         \\
                  & max\_features: [``auto'', ``sqrt'']                \\
                  & n\_estimators: [10, 20, 50, 100, 200]              \\
    Ridge         & alpha: $[1e^{-10},1e^{-9},\dotsc,30.0]$            \\
    SVR           & C: [0.01, 1.34, 2.67, 4.0]                         \\
                  & kernel = [``linear'', ``poly'', ``rbf'']           \\
    \bottomrule
\end{tabular}
\end{small}
\caption{Hyper-parameter search space.}\label{tab:hyperparameters}
\end{table}

\textbf{Results}: We collected three different metrics to have better insight into the regression loss errors in the trained models mentioned above: the Mean Absolute Percentage Error (MAPE), the Mean Squared Errors (MSE), and the Mean Absolute Error (MAE). Despite using a multiple scoring strategy to evaluate the CV on the test set, it was necessary to choose a single scorer to refit the estimators on the whole dataset using the best-found parameters. For that purpose, we selected the MAPE.

\cref{tab:training_and_validation_errors} shows the result of the hyper-parameter search with the three aforementioned metrics. The table highlights the minimum error for each metric in training, validating, and testing. From the results presented, we were able to confirm that \textbf{RQ1} is correct and the feature vector is good at characterizing the performance of the reservoir simulator. The selected regressor, Random Forest, was capable of making predictions with a 6.2\%, 8.0\%, and 16.3\% MAPE, in train, validation, and test, respectively. As we used the LOGO-CV strategy, we can be assured that \textbf{RQ2} was also fulfilled as the test errors are very reasonable, despite being slightly higher, as expected for an unseen model.

From~\cref{tab:training_and_validation_errors} we can see that the Random Forest Regressor achieved the best results with almost all metrics (the lower the better). For the cases where it does not have the lowest error, we can see that it is not far from the minimum achieved. For that reason, we selected this trained model to be the Oracle used in the experiments to answer \textbf{RQ3} and \textbf{RQ4} in ~\cref{sec:experiments:performance-model-validation}.

\begin{table*}[!htbp]
    \begin{center}
    \begin{small}
    \begin{tabular}{@{}lcrrrcrrrcrrr@{}}
    \toprule
    {Metric}                  & \phantom{} & \multicolumn{3}{c}{MAPE (\%)}       & \phantom{} & \multicolumn{3}{c}{MSE (seconds$^2$)}         & \phantom{} & \multicolumn{3}{c}{MAE (seconds)}       \\
                                             \cmidrule{3-5}                                     \cmidrule{7-9}                                               \cmidrule{11-13}                        \\
    {}                        &            &     train & validation & test       &            &         train & validation    & test          &            &       train &  validation & test        \\
    Regression Model          &            &           &            &            &            &               &               &               &            &             &             &             \\
    \midrule
    Decision Tree             &            &     16.4  &      17.2  &      18.1  &            & {\bf 9.8e+05} &      1.4e+07  & {\bf 1.2e+07} &            &       346.0 &       636.0 & {\bf 649.2} \\
    Extremely Randomized Tree &            &  1,533.8  &   1,547.2  &     196.6  &            &      1.4e+08  &      1.4e+08  &      1.2e+08  &            &     4,373.3 &     4,413.5 &     4,543.1 \\
    Elastic Net               &            &     12.5  &      15.4  &      23.1  &            &      1.2e+06  &      2.1e+07  &      1.4e+07  &            &       245.9 &       645,3 &       757.1 \\
    Gradient Boosting         &            &    412.2  &     414.5  &      68.9  &            &      4.6e+07  &      5.4e+07  &      1.6e+08  &            &     1,303.8 &     1,520.0 &     3,869.7 \\
    K-nearest Neighbors       &            &      8.7  &      15.8  & {\bf 15.1} &            &      1.3e+07  &      3.1e+07  &      4.1e+07  &            &       441.6 &       750.5 &       886.6 \\
    Lasso                     &            &  1,538.4  &   1,533.6  &     274.0  &            &      1.1e+08  &      1.2e+08  &      1.1e+08  &            &     4,153.4 &     4,199.4 &     4,172.3 \\
    Multi-layer Perceptron    &            &  2,695.5  &   2,720.8  &     219.9  &            &      4.3e+08  &      4.4e+08  &      2.7e+08  &            &     8,213,7 &     8,326.4 &     9,052.7 \\
    Random Forest             &            & {\bf 6.2} &  {\bf 8.0} &      16.3  &            &      1.9e+06  & {\bf 9.8e+06} &      1.7e+07  &            & {\bf 220.7} & {\bf 434.2} &       665.8 \\
    Ridge                     &            &  1,566.4  &   1,582.8  &     232.7  &            &      1.1e+08  &      1.2e+08  &      1.1e+08  &            &     4,157.9 &     4,205.6 &     4,258.4 \\
    Linear Support Vector     &            &    259.0  &    259.3   &     897.7  &            &      8.6e+08  &      8.6e+08  &      9.1e+08  &            &     5,713.1 &     5,714.9 &     6,094.1 \\
    \bottomrule
    \end{tabular}
    \caption{Training, validation and test errors for the selected regressor models with different metrics.}
    \label{tab:training_and_validation_errors}
    \end{small}
    \end{center}
    \end{table*}

\subsection{Performance Model Validation}\label{sec:experiments:performance-model-validation}

To validate \textbf{RQ3} and \textbf{RQ4} we executed \cref{alg:pseudocode} with the three models detailed in \cref{tab:workloads_validation} that were not seen during the training of the performance model. The plots in this section represent the results of running our implementation of the ES-MDA with and without TunaOil (commenting lines \ref{lst:line:tuna-begin} through \ref{lst:line:tuna-end}). We used the Random Forest Regressor in all tests with TunaOil. We also compared the performance and quality results of TunaOil against the default settings of the simulator and the pre-defined engineer numerical settings.

\begin{figure*}[tb]
\centering
\begin{subfigure}[b]{0.475\textwidth}
    \centering
    \includegraphics[scale=0.5]{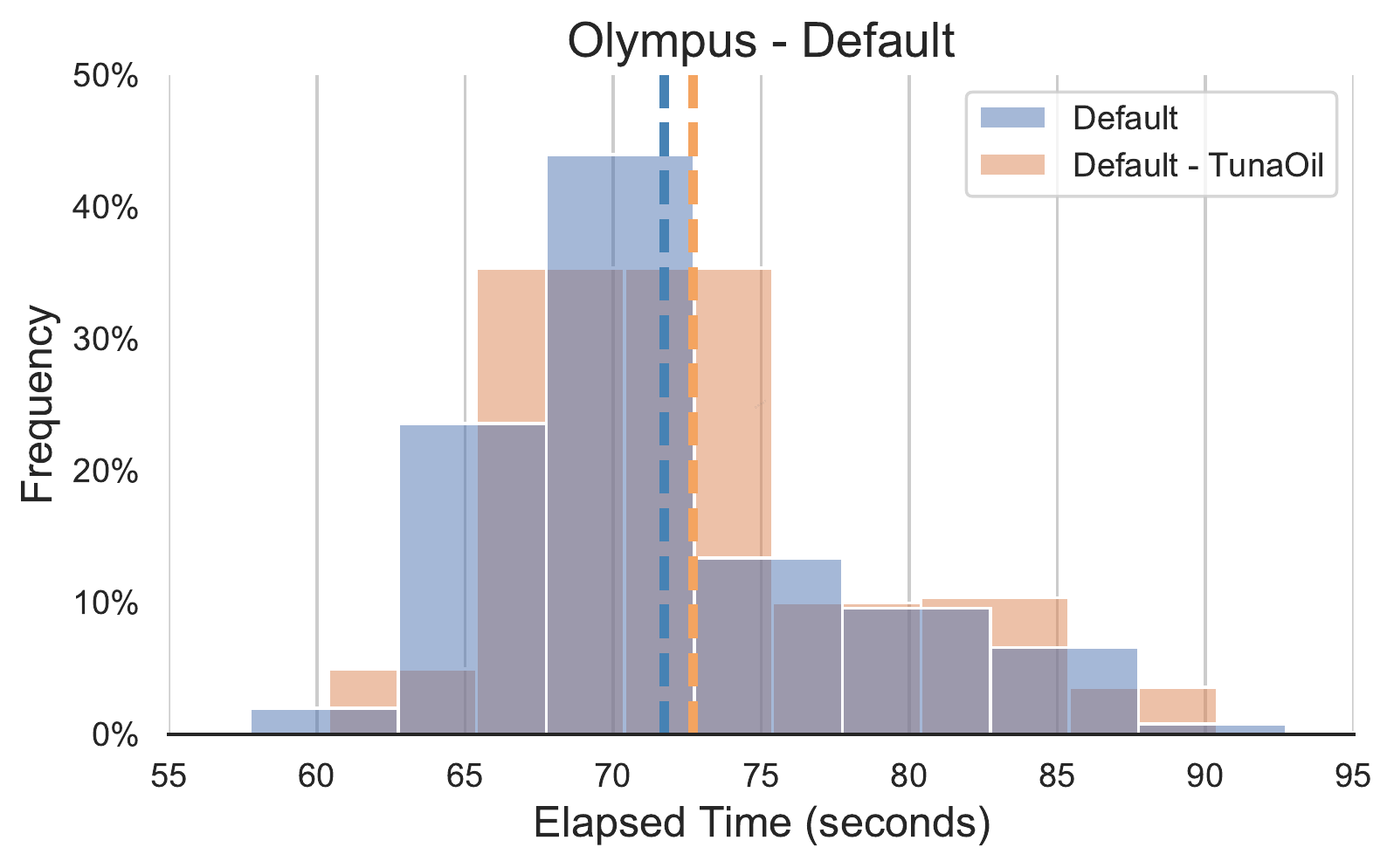}
    \caption{Elapsed time with the default settings.}
    \label{fig:results_olympus_default_elapsed_time}
\end{subfigure}
\hfill
\begin{subfigure}[b]{0.475\textwidth}
    \centering
    \includegraphics[scale=0.5]{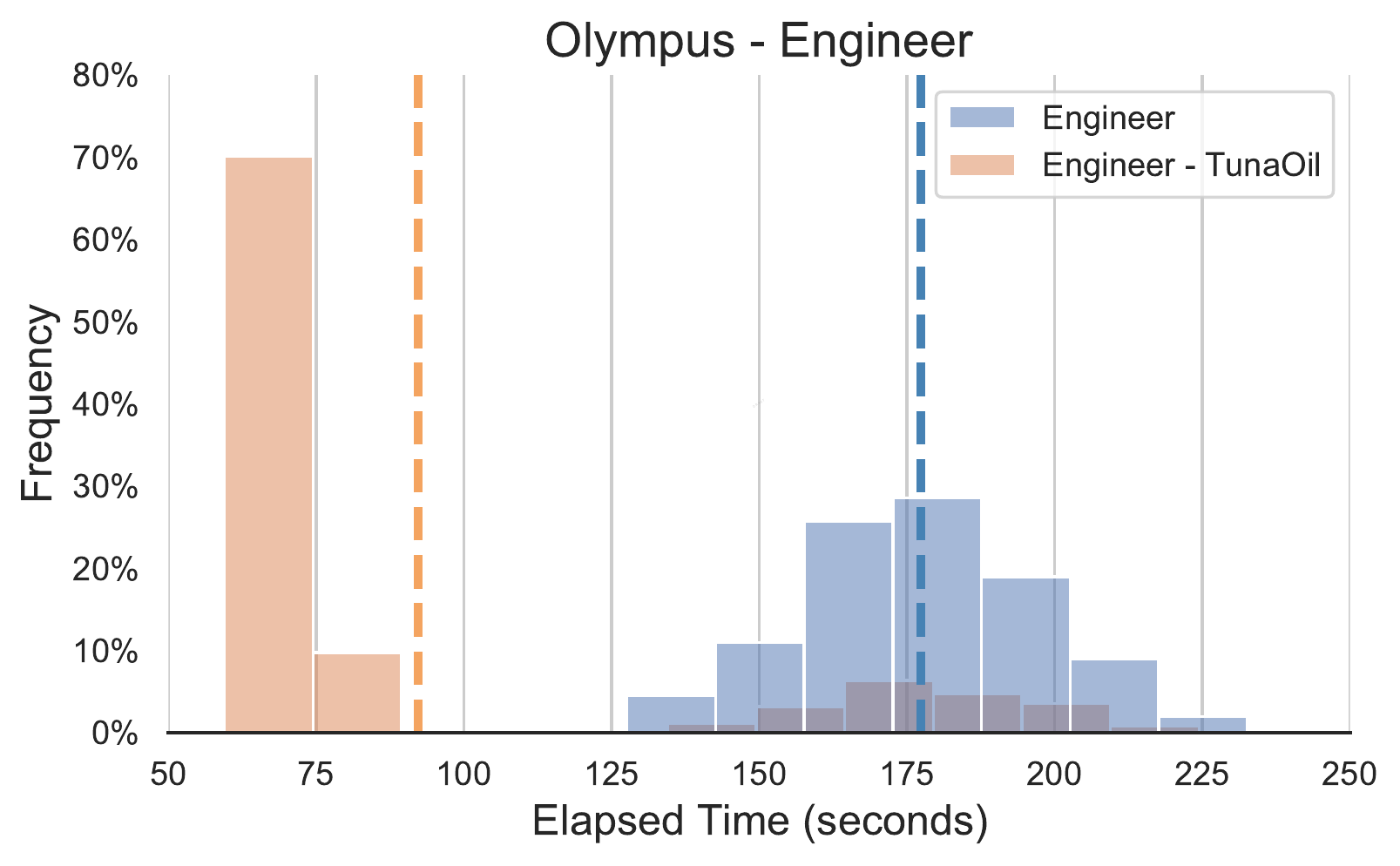}
    \caption{Elapsed time with the engineer numerical settings.}
    \label{fig:results_olympus_engineer_elapsed_time}
\end{subfigure}
\vfill\vspace{0.8cm}
\begin{subfigure}[b]{0.475\textwidth}
    \centering
    \includegraphics[scale=0.5]{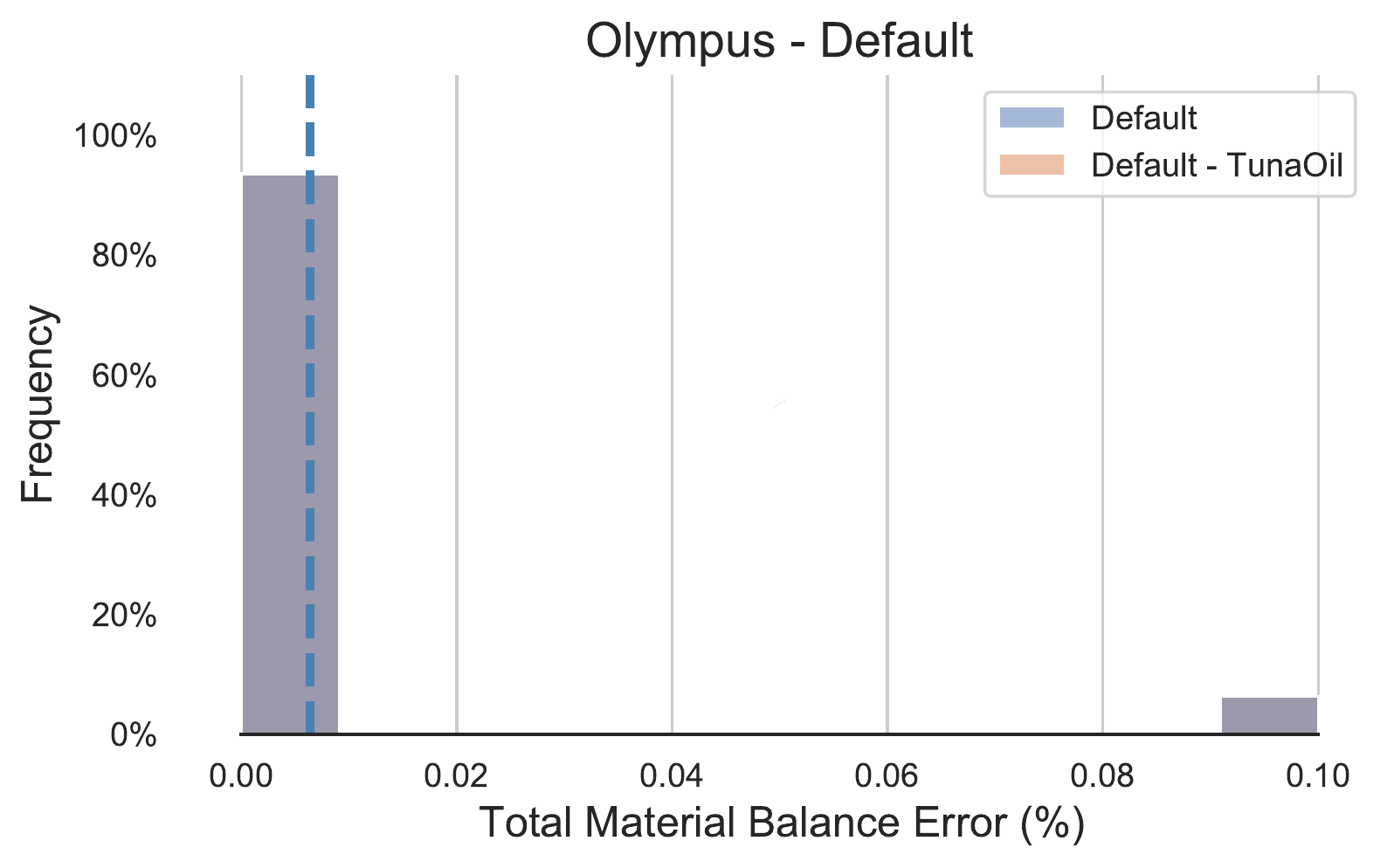}
    \caption{Total MBE with the default settings.}
    \label{fig:results_olympus_default_mbe}
\end{subfigure}
\hfill
\begin{subfigure}[b]{0.475\textwidth}
    \centering
    \includegraphics[scale=0.5]{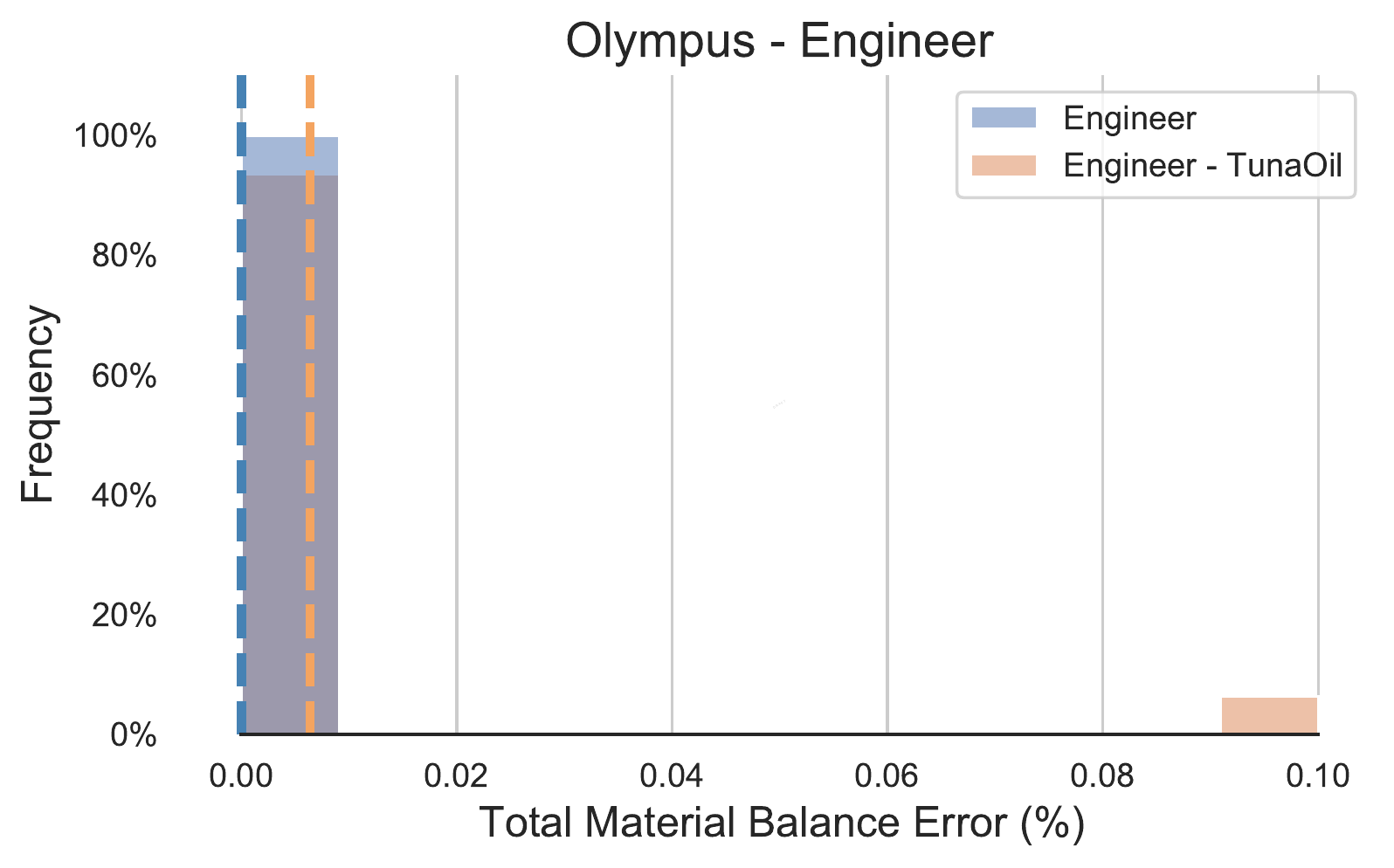}
    \caption{Total MBE with the engineer settings.}
    \label{fig:results_olympus_engineer_mbe}
\end{subfigure}
\caption{OLYMPUS histograms for the 250 cases executed in an ES-MDA with and without TunaOil. The dashed lines represent the mean of the values.}
\label{fig:results_olympus}
\end{figure*}

\begin{figure*}[tb]
\hfill
\begin{subfigure}[tb]{0.3\textwidth}
\centering
\includegraphics[scale=0.6,trim=80 150 80 0,clip]{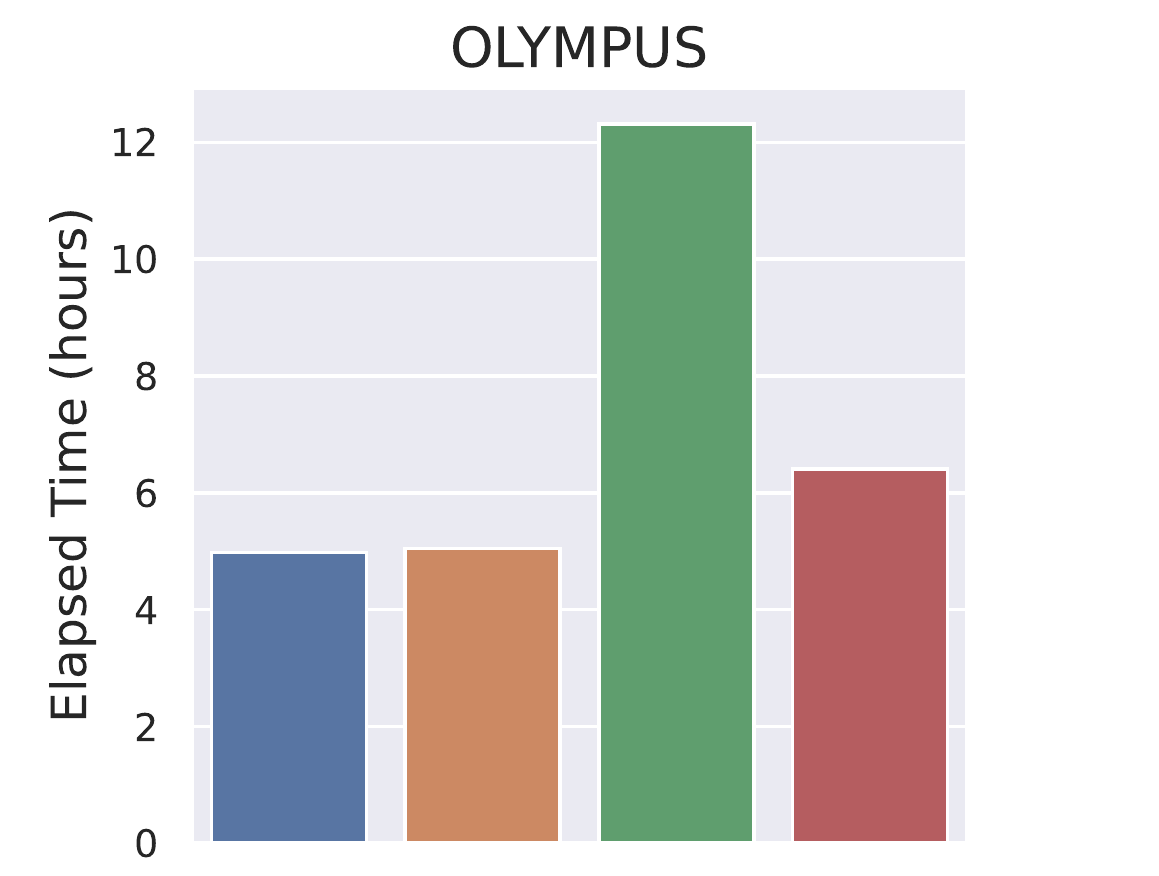}
\caption{}%
\label{fig:results_olympus_total_times}
\end{subfigure}
\hfill
\begin{subfigure}[tb]{0.3\textwidth}
\centering
\includegraphics[scale=0.6,trim=80 150 80 0,clip]{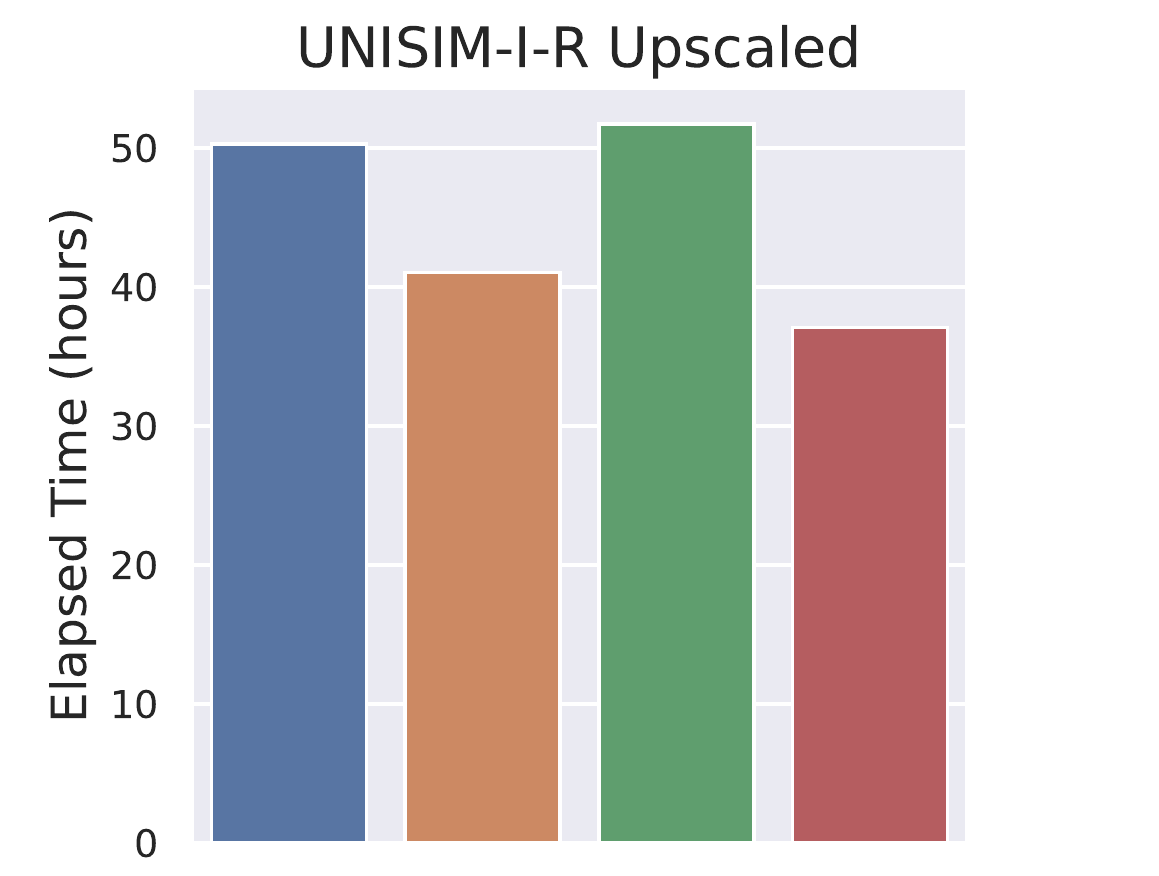}  %
\caption{}%
\label{fig:results_unisim-i_upscaled_total_times}
\end{subfigure}
\hfill
\begin{subfigure}[tb]{0.3\textwidth}
\centering
\includegraphics[scale=0.6,trim=80 150 80 0,clip]{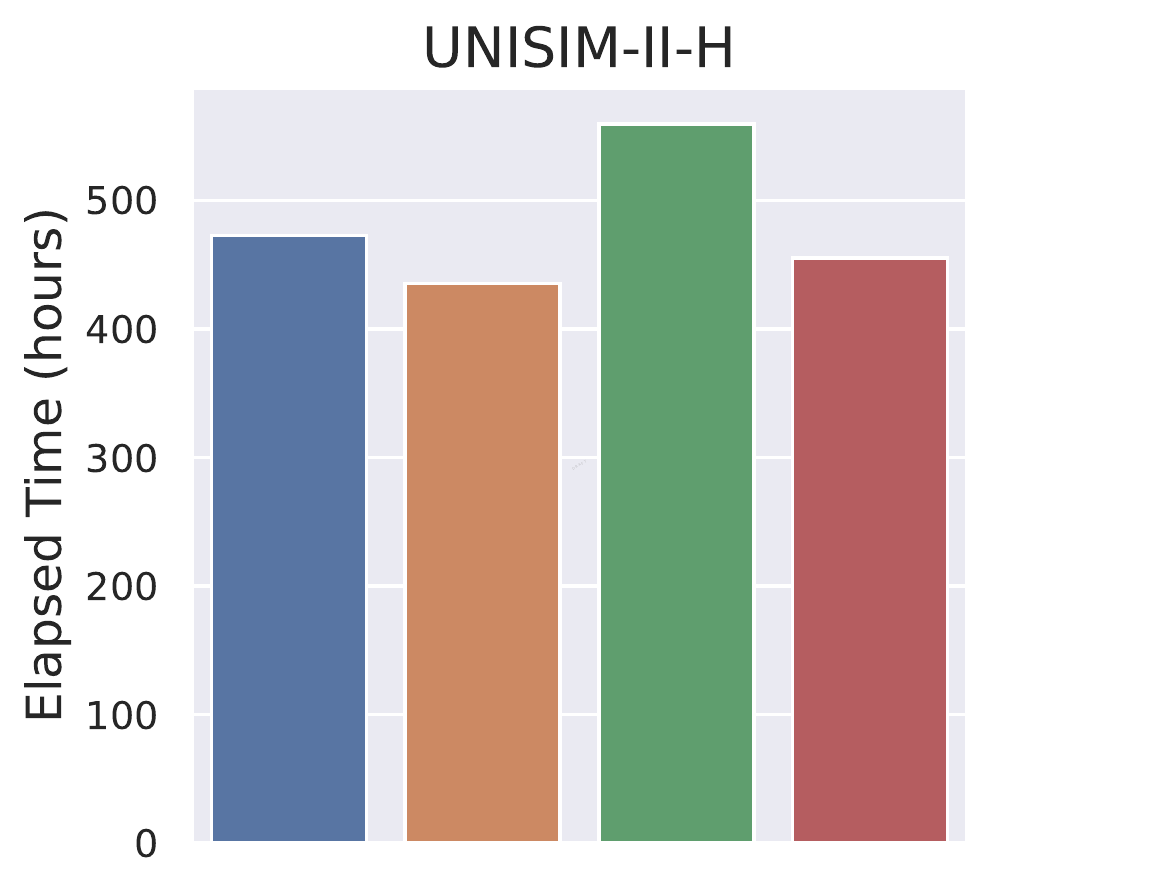}
\caption{}%
\label{fig:results_unisim-ii_total_times}
\end{subfigure}
\hfill
\par
\begin{subfigure}[tb]{\textwidth}
\centering
\includegraphics[scale=0.6,trim=0 0 0 360,clip]{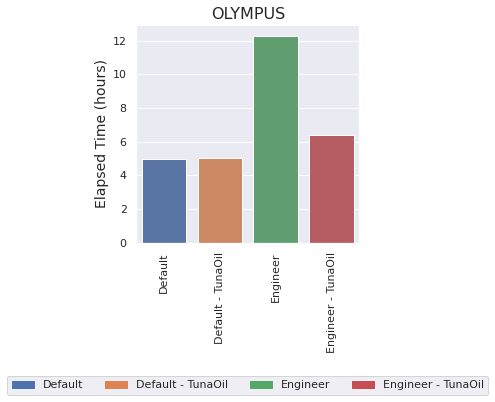}
\end{subfigure}
\caption{Total elapsed time of all ES-MDA simulations in each of the four scenarios for the three validation workloads.
}
\label{fig:results_total_times}
\end{figure*}

\cref{fig:results_olympus} shows histograms for the \textbf{OLYMPUS} workload with it 250 simulations ES-MDA, comparing the usage of TunaOil with two baselines: the default and the engineer parameters. The engineer case speedup was $1.92\times$, with a marginal increase in MBE, as shown in~\cref{fig:results_olympus_engineer_mbe}. However, for the default baseline, TunaOil was unable to improve the elapsed total time. Nevertheless, TunaOil is very close to the default performance, being only 1\% slower and with the same MBE quality. It is interesting to observe in \cref{fig:results_olympus_total_times} that the default parameters perform better than the parameters from the engineer. As briefly discussed in~\cref{sec:workloads:validation-models}, this could be an effect of the conversion between simulators, where the engineer chose to tighten the numerical controls significantly to make the result and quality as close as possible to the original model in Eclipse format.
This was the simplest workload from the three chosen for the validation, as it uses a simple \texttt{OIL-WATER} model, and so it is understandable that TunaOil performed poorly against the Default parameters as it did not have many numerical opportunities to explore.

\begin{figure*}[tb]
\centering
\begin{subfigure}[b]{0.475\textwidth}
    \centering
    \includegraphics[scale=0.55]{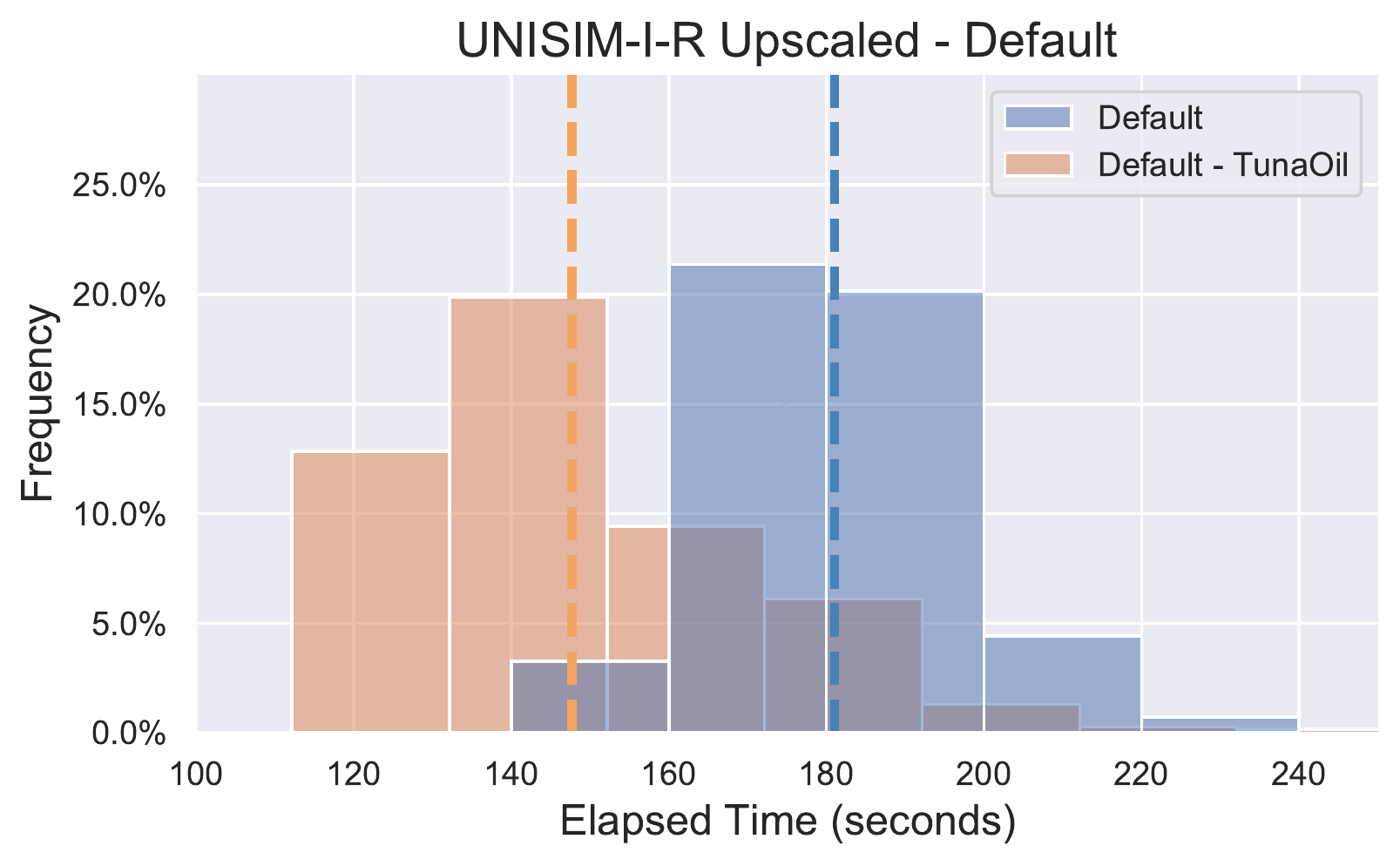}
    \caption{Elapsed time with the default settings.}
    \label{fig:results_unisim-i_upscaled_default_elapsed_time}
\end{subfigure}
\hfill
\begin{subfigure}[b]{0.475\textwidth}
    \centering
    \includegraphics[scale=0.55]{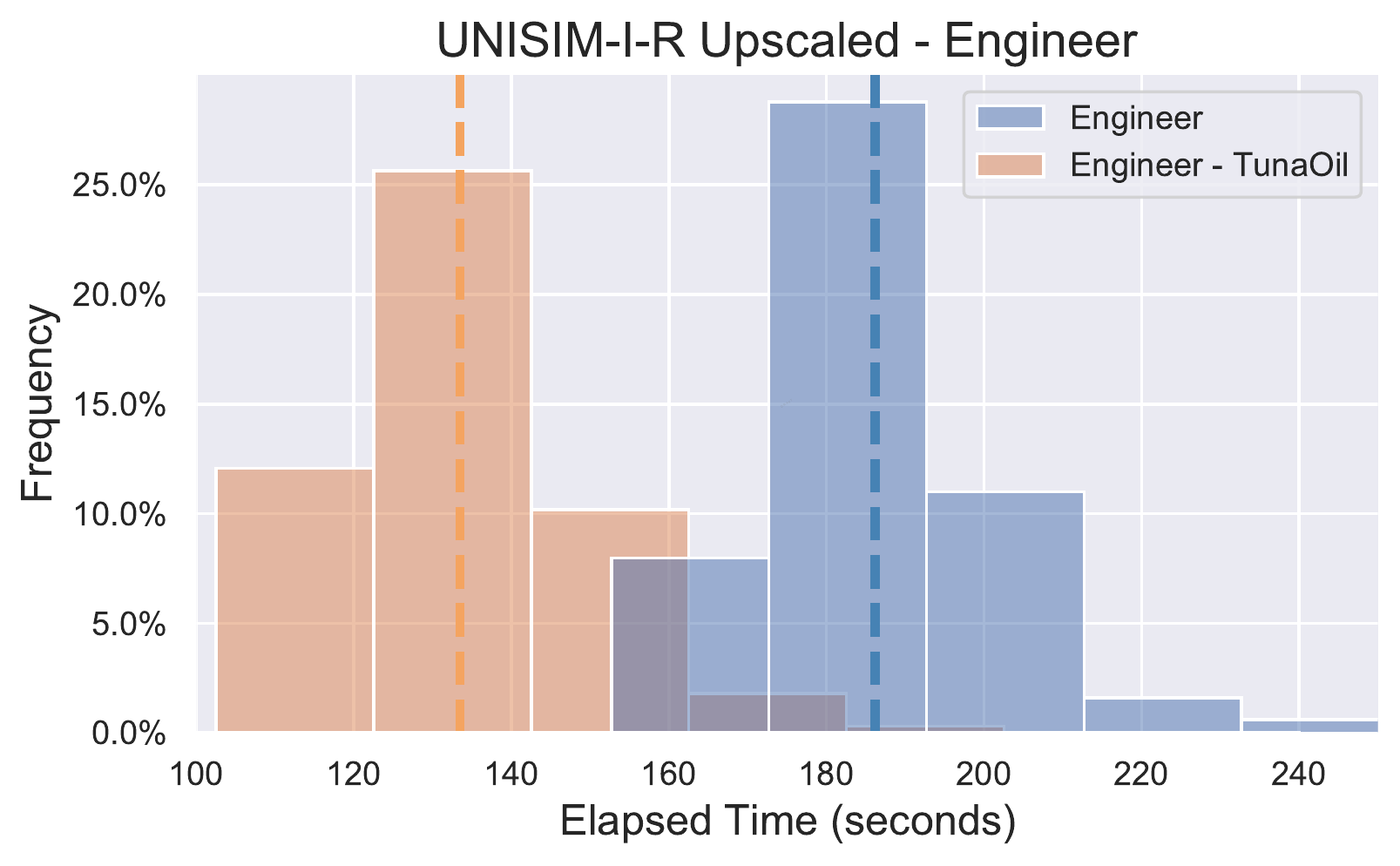}
    \caption{Elapsed time with the engineer numerical settings.}
    \label{fig:results_unisim-i_upscaled_engineer_elapsed_time}
\end{subfigure}
\vfill
\begin{subfigure}[b]{0.475\textwidth}
    \centering
    \includegraphics[scale=0.55]{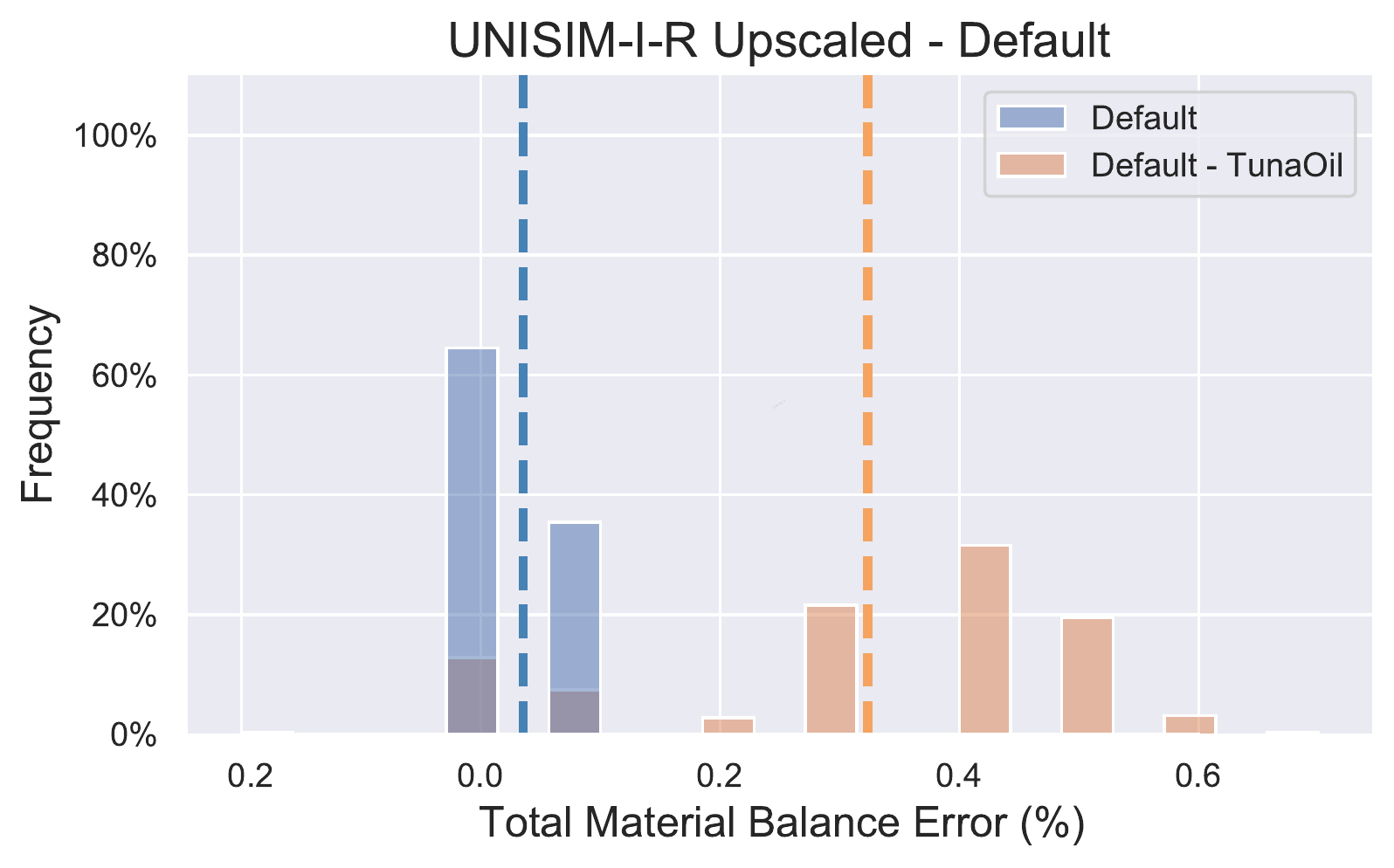}
    \caption{Total MBE with the default settings.}
    \label{fig:results_unisim-i_upscaled_default_mbe}
\end{subfigure}
\hfill
\begin{subfigure}[b]{0.475\textwidth}
    \centering
    \includegraphics[scale=0.55]{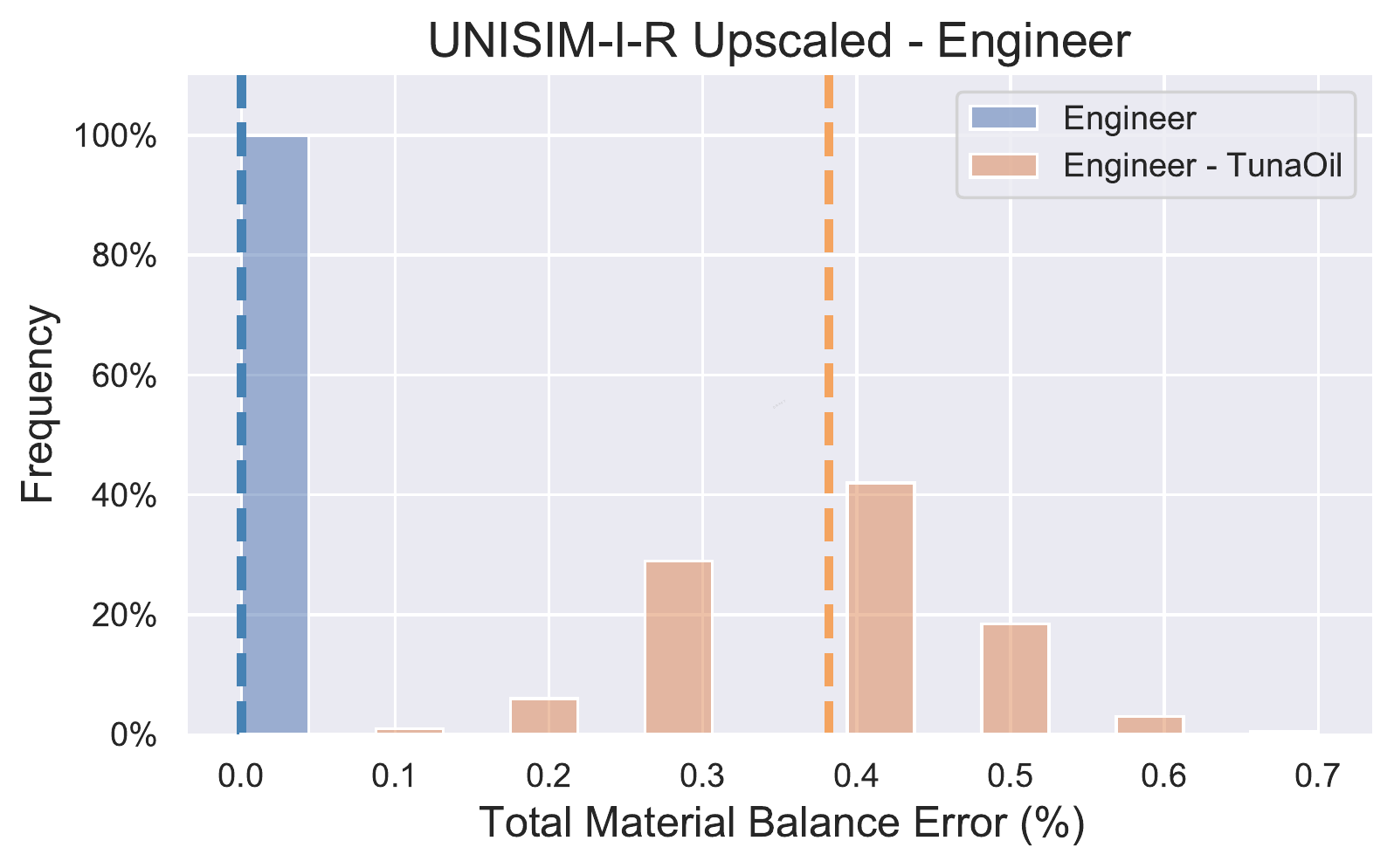}
    \caption{Total MBE with the engineer settings.}
    \label{fig:results_unisim-i_upscaled_engineer_mbe}
\end{subfigure}
\caption{UNISIM-I histograms for the 240 cases executed in an ES-MDA with and without TunaOil. The dashed lines represent the mean of the values.}
\label{fig:results_unisim-i_upscaled}
\end{figure*}

\cref{fig:results_unisim-i_upscaled} shows similar histograms with the distribution of all the 240 simulations each ES-MDA performed for the \textbf{UNISIM-I-R Upscaled} HM case. The results are split in terms of total elapsed time of the simulations using the default and engineer numerical values as baseline, \ref{fig:results_unisim-i_upscaled_default_elapsed_time} and \ref{fig:results_unisim-i_upscaled_engineer_elapsed_time} respectively, and the MBE, \ref{fig:results_unisim-i_upscaled_default_mbe} and \ref{fig:results_unisim-i_upscaled_engineer_mbe} respectively, representing the quality metric evaluated. TunaOil provided a speedup of $1.23\times$ regarding the default numerical parameters from the simulator and $1.39\times$ with the parameters available in the original UNISIM model. The speedup was calculated considering the mean time of all simulations (the dashed lines values in Figures \ref{fig:results_unisim-i_upscaled_default_elapsed_time} and \ref{fig:results_unisim-i_upscaled_engineer_elapsed_time}), and it is equivalent to the total elapsed time speedup seen in \cref{fig:results_unisim-i_upscaled_total_times}. These speed ups were the consequence of more relaxed parameters being used in the simulations that led to some simulations having a slightly higher MBE. Despite the increase in the error, it is always below 1\%, which is very acceptable from the reservoir engineering perspective.

\begin{figure*}[tb]
\centering
\begin{subfigure}[b]{0.475\textwidth}
    \centering
    \includegraphics[scale=0.55]{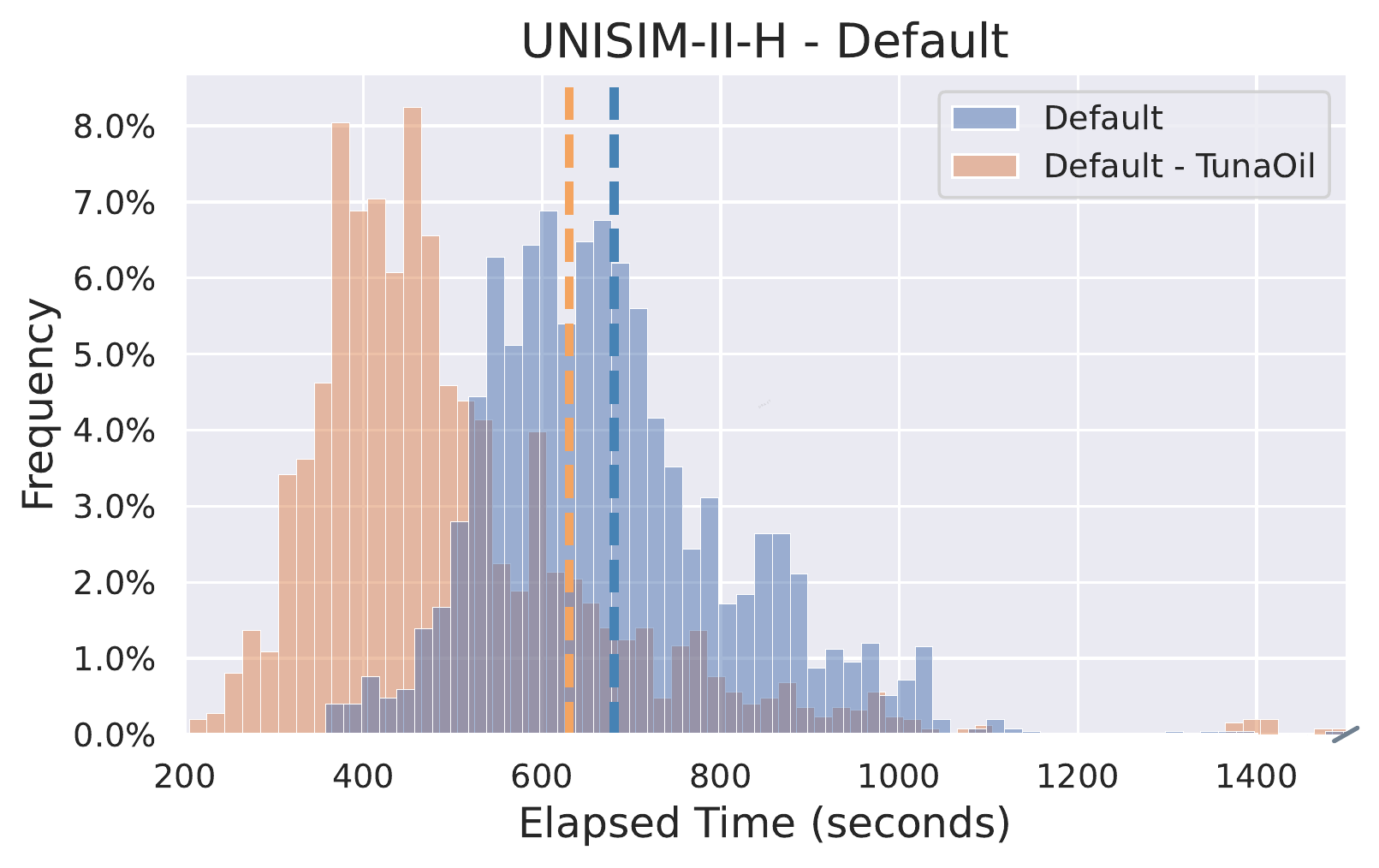}
    \caption{Elapsed time with the default settings.}
    \label{fig:results_unisim-ii_default_elapsed_time}
\end{subfigure}
\hfill
\begin{subfigure}[b]{0.475\textwidth}
    \centering
    \includegraphics[scale=0.55]{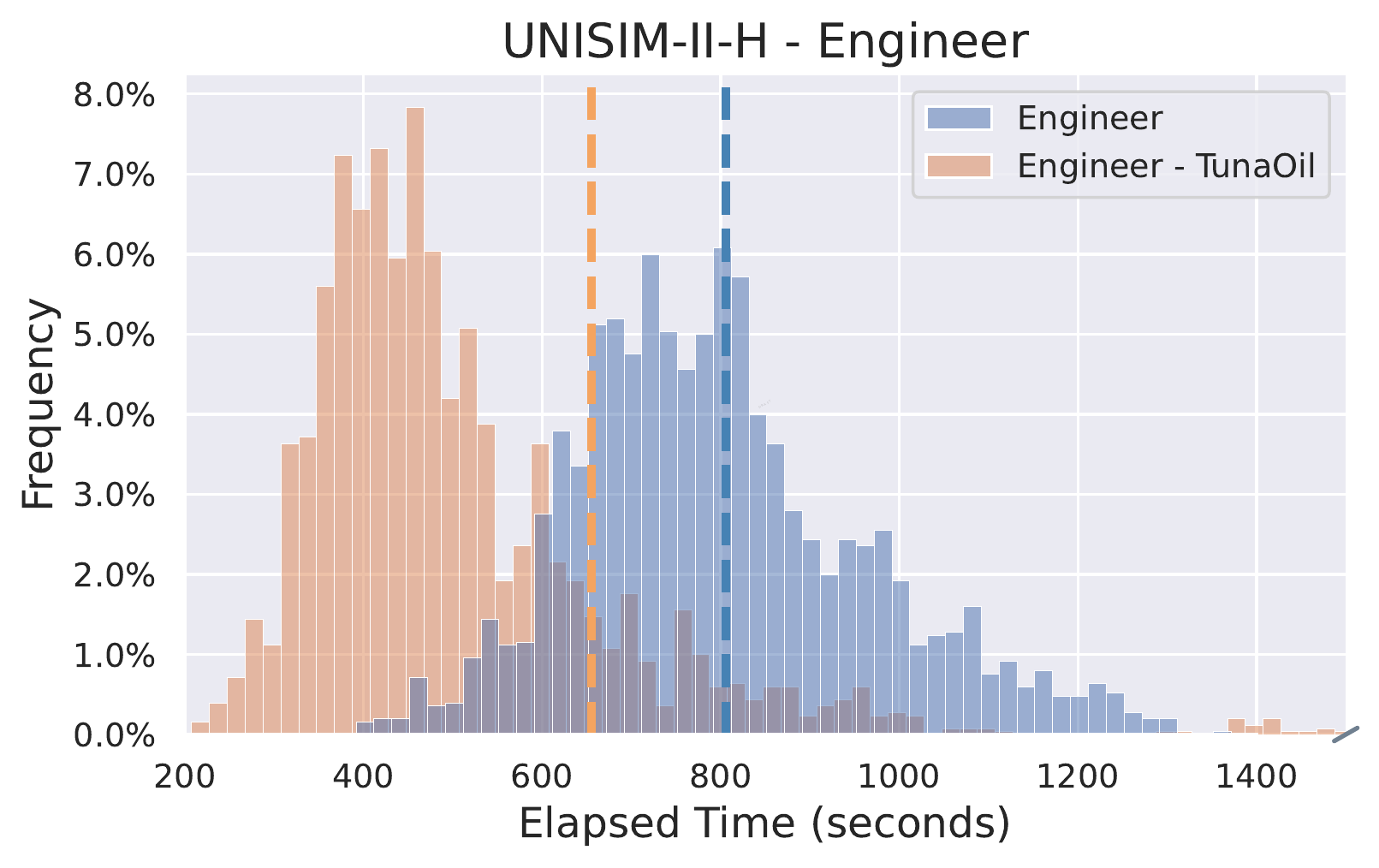}
    \caption{Elapsed time with the engineer numerical settings.}
    \label{fig:results_unisim-ii_engineer_elapsed_time}
\end{subfigure}
\vfill
\begin{subfigure}[b]{0.475\textwidth}
    \centering
    \includegraphics[scale=0.55]{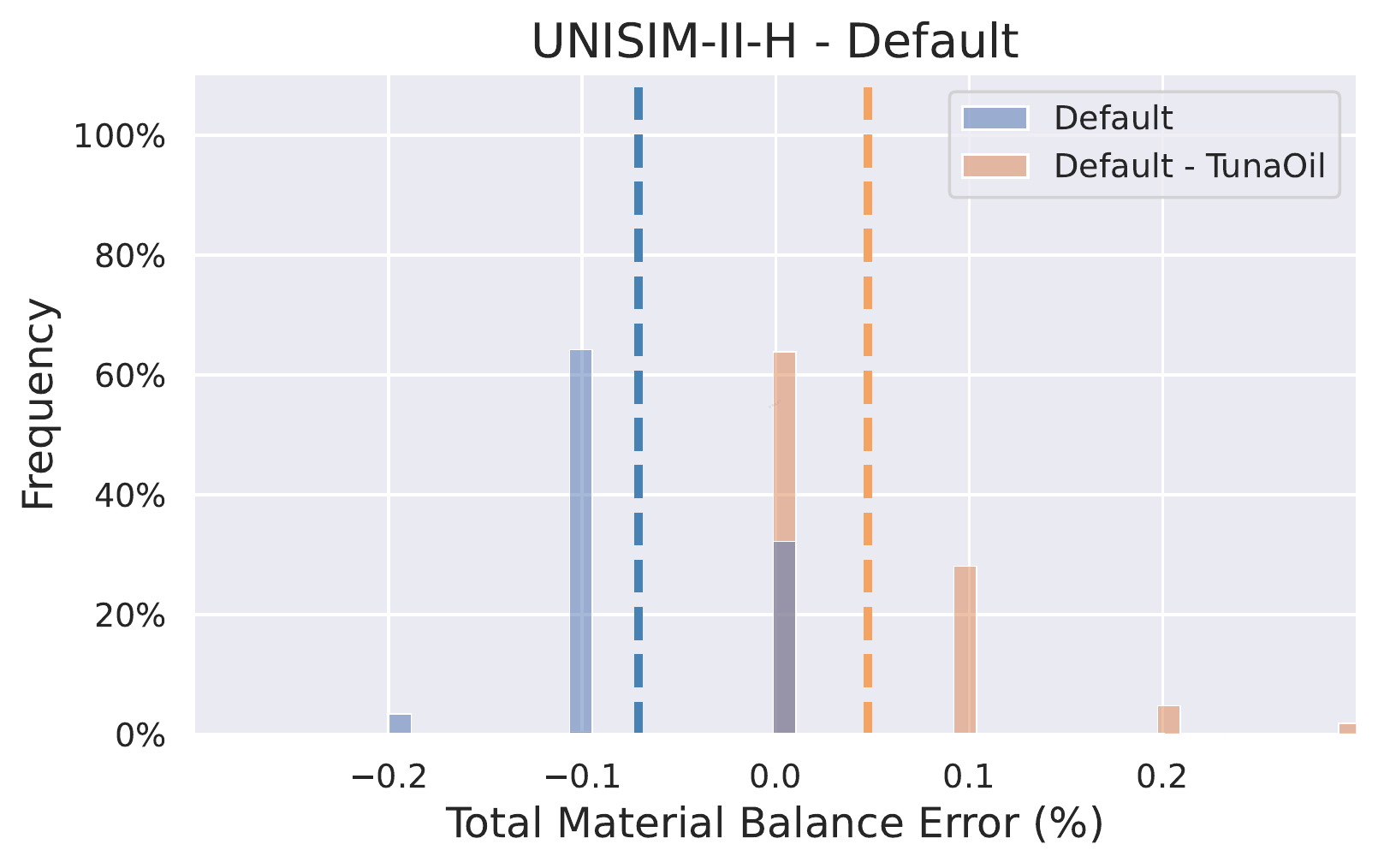}
    \caption{Total MBE with the default settings.}
    \label{fig:results_unisim-ii_default_mbe}
\end{subfigure}
\hfill
\begin{subfigure}[b]{0.475\textwidth}
    \centering
    \includegraphics[scale=0.55]{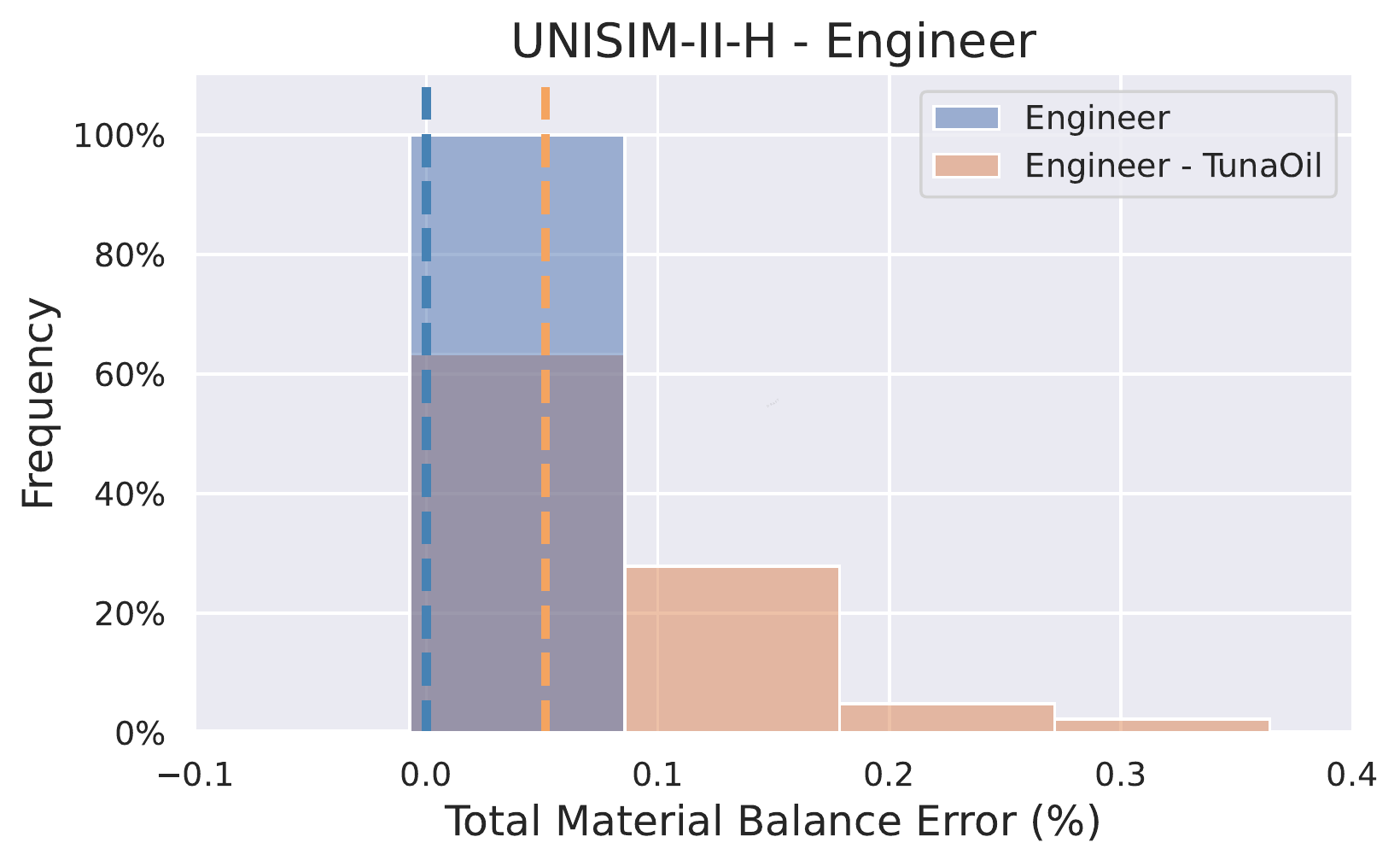}
    \caption{Total MBE with the engineer settings.}
    \label{fig:results_unisim-ii_engineer_mbe}
\end{subfigure}
\caption{UNISIM-II histograms for the 500 cases executed in an ES-MDA with and without TunaOil. The left-hand side gives the plots considering the Default parameters as the baseline, and the right-hand side, the plots comparing with the Engineer configurations. The upper row gives the performance plots with the histogram of the elapsed time. In the bottom row, the quality is represented by the total material balance error histogram. The dashed lines represent the mean of the values.}
\label{fig:results_unisim-ii}
\end{figure*}

The last experiment, shown in \cref{fig:results_unisim-ii}, is for \textbf{UNISIM-II}. This case requires 2,500 simulations for each ES-MDA analysis performed as it has 500 realizations. Therefore, a given configuration was only executed once, unlike with OLYMPUS and UNISIM-I, which we executed three times and plotted the mean. TunaOil was 7.98\% faster than the Default parameter set and 22.87\% faster when compared with the Engineer parameters. TunaOil reduced the mean elapsed time from 680.89 to 630.55 seconds and from 805.56 to 655.60 seconds for the Default and Engineer configurations, respectively.

Regarding the quality of the simulations, we can see from Figures \ref{fig:results_unisim-ii_engineer_mbe} and \ref{fig:results_unisim-ii_default_mbe} that the results of TunaOil were similar to previous experiments, with the slight changes to the MBE being reasonable and within very acceptable ranges from the engineering perspective. The mean MBE of the ES-MDA with Defaults, in fact, is marginally better as it is closer to zero with TunaOil (0.048) rather than without it (-0.071). From an alternative point of view, we have a higher percentage of cases with 0.0 MBE with TunaOil, so in this sense, TunaOil was capable of providing faster and better simulations.

Contrary to the previous experiments with OLYMPUS and UNISIM-I,  TunaOil struggles to improve some realizations of UNISIM-II, which might be a consequence of the increased complexity in the physics of the model discussed in \cref{sec:workloads:validation-models}. Therefore, Figures \ref{fig:results_unisim-ii_default_elapsed_time} and \ref{fig:results_unisim-ii_engineer_elapsed_time} has the x-axis truncated at 1,500 seconds for proper visualization of the main distribution of the elapsed time histogram. On the extreme right-hand side of those plots, we can see a few realizations that already took longer to run. Furthermore, three realizations (174, 322, and 382) are completely off the axis, with more than 10,000 seconds, in all the simulations with TunaOil parameters (iterations 2 to 5 of the ES-MDA). An analysis of the simulation log of those cases reveals that the pattern was the same. In the last year of the simulation (somewhere around 2025), they start to face convergence problems and cut the timesteps from days to hours then to just a few minutes, which leads, in the worst case of realization 382, to it taking almost 10 hours (34.838 seconds) to complete the simulation. The NI/TS was 19.13 and the LI/NI only 1.04, for this problematic case. A similar pattern of dozens of Newton Iterations per timestep with almost a single Solver Iteration was found in all those slow realizations. In the cases that had a good performance, such as realization 409, we can observe an NI/TS of 2.02 with an LI/TS of 3.15.

The aggregated simulation time for each ES-MDA study performed (with and without TunaOil for both baseline parameters) is given in~\cref{fig:results_total_times}. For the \textbf{OLYMPUS} workload, we can observe in ~\cref{fig:results_olympus_total_times} that the engineer case speedup was $1.92\times$, reducing the overall HM time from 12:18:50 hours to 6:24:26 hours. For the default the total ES-MDA time increased from 04:58:52 hours to 05:02:54 hours, a slow down of $0.99{\times}$.
In the \textbf{UNISIM-I-R Upscaled} workload, the total elapsed time with the default parameters was reduced from 50:16:53 hours to 41:01:17 hours, a speedup of $1.23\times$. The cases with the engineer parameters were reduced from 51:42:32 to 37:04:58, a speedup of $1.39\times$ as shown in ~\cref{fig:results_unisim-i_upscaled_total_times}. For the \textbf{UNISIM-II} workload, we observed a reduction of almost 38 hours, from 472:50:24 hours to 436:15:14 hours with the Default configurations, which represents a speedup of $1.08{\times}$. Considering the Engineer baseline, the reduction was from 559:24:54 to 455:16:48 hours, a reduction of 104-hours, which represents a speedup of $1.23{\times}$. On average, the speed up among all workloads and configurations was $1.31{\times}$. If we compare using the Default numerical parameter of the simulator or the tuned one by the Engineer, we have speedups of $1.10{\times}$ and $1.52{\times}$, respectively.

\subsection{Computational Environment}
\label{sec:experiment:computational_environment}

All the experiments in this work were performed in two HPC environments: Santos Dumont and BSCDC.

\begin{itemize}
    \item The first, also know as SDumont, is a 5.1 PFlops supercomputer from LNCC\footnote{\MYhref{https://sdumont.lncc.br/}{https://sdumont.lncc.br/}}, which was used to execute all the reservoir simulations. The supercomputer has a mixed node configuration, but the partition used consisted of 246 CPU nodes, each with 2x Intel Xeon Cascade Lake Gold 6252 (24 physical cores per socket) and 384 Gb of memory. The storage equipment for the scratch is an HPE ClusterStor 9000 using an integrated Lustre file system. All the simulations were fixed to use the 48 physical cores available.
    \item The second cluster is one infrastructure from the Data-Centric Computing Group of BSC\footnote{BSC Data-Centric Computing group \MYhref{https://www.bsc.es/discover-bsc/organisation/scientific-structure/data-centric-computing}{web page link}}. It has a hybrid node configuration, but the nodes used all have an 2x Intel Xeon Broadwell E5-2630 v4 processor with 128 GB of RAM. This cluster was used for the development and testing of the code, machine learning training, hyper-parameter optimization, etc.
\end{itemize}

The black-oil reservoir simulator used was the CMG IMEX\footnote{\MYhref{https://www.cmgl.ca/imex}{https://www.cmgl.ca/imex}} version 2019.10. The code of the experiments used Python 3.x with scikit-learn as the main machine learning library. We used the LHS and the BO implementations from the scikit-optimize library. The ES-MDA code was implemented without localization support and without resilience support on failing simulations.

\section{Conclusions}\label{sec:conclusion}

This paper introduces TunaOil, a performance model methodology used to tune the numerical parameters of black-oil reservoir models within a history matching (HM) workflow. In the worst case, TunaOil presented only a 1\% slower performance for the HM study than using the default configuration of the simulator. Though for most of the workloads evaluated, TunaOil reduced the overall HM runtime, on average by 31\%, without demanding additional simulations or a separated optimization study. An enhancement of 10\% when compared with the Default parameters and a 52\% when comparing with the Engineer without major impact in the quality. Our experiments demonstrated that the Oracle built was able to predict the proper effect of the changes in the solver options in terms of simulation time and quality (maintaining the accepted engineering quality of the studies). The experiments have shown that our Oracle makes accurate predictions (16.3\% of mean absolute percentage error in the test dataset) in a broadly used workflow in the petroleum engineering area with black-oil models, but the idea can be easily extended to other types of workflows, such as optimizations processes (for well placement, well controls, etc.) or other types of models, such as compositional models. However, other features may need to be considered, e.g., the number of components, as this fluid property can impact deeply on the performance of compositional simulators. Ultimately, we could couple the Oracle in the central scheduler system of an energy company to perform a live optimization of any reservoir simulation being submitted to an HPC infrastructure (on-premises or on a cloud provider), reducing time to solution and associated costs.

The premises of our work include fixing the number of cores that were used by the simulator. Typically, for a reservoir model with millions of cells and simulated exclusively in a single compute node, we must use all CPU cores to achieve the minimum simulation execution time. The scenario is different when we deal with small/medium models or considering the collocation of jobs (side-by-side jobs on the same node). In a future study, we want to include the number of threads (PNTHREADS) in the parameter space. In fact, we did a quick evaluation of this approach, and it brings new challenges as the parallelism has a huge impact on the performance of the simulator and has a direct effect on the other parameters. A good set of parameters for 48 cores may not represent a suitable configuration for a different number of process threads. 
Our preliminary evaluation has shown better results might be achieved by splitting the Oracle into a classification for the best PNTHREADS parameter and then performing the regression for the best numerical parameters. TunaOil, as presented in this work, focuses only on the latter part.
Furthermore, the architecture of the machine being used has a direct influence on this classification system. The recommendation of the best hardware, which can include the best processor to minimize the simulation time or the minimum memory required, could be included in the predictors. 
Another relevant future work is to evaluate better the importance of each feature used in the Oracle. We can even consider additional features such as the perforation interval or per time-step properties (e.g., the number of Newton iterations per time-step), as we only considered aggregated statistics of the simulation. We included a Feature Selection step in the pipeline, but we did not dig into a more meticulous analysis of the importance of each one to the performance of TunaOil. We can analyze them better with the baseline of this work and try to archive the same performance results with fewer features, speeding up the training/prediction steps, or even archieving better performance by aggregating new features.
Apart from improving the quality of the Oracle, there is room for other potential search techniques to query the Oracle built and find better ``optimal'' values for each realization. Even the SBO that we explored initially could be optimized to query the multiple realizations in parallel, reducing the computational overhead and improving the speedup/quality when dealing with multiple realization scenarios.

Although TunaOil generated a few problematic simulations in terms of performance for the UNISIM-II workload, the overall ES-MDA was improved against both baselines - Defaults and Engineer. In any case, some ideas could be explored to strengthen those results by enhancing the regressor or limiting the impact of a bad configuration. From the ML perspective, we could have a previous classification step to cherry pick the numerical parameters that should be explored in the query phase. From an architecture perspective, we could apply a timeout limit to all the simulations being generated by TunaOil, avoiding any spurious long simulations even when the suggested numerical parameters are not enough to prevent that. This would require a more resilient ES-MDA implementation to handle some realizations' failures, which is a good approach in general as the simulations could fail for other reasons: from hardware failures of the computer nodes to numerical convergence problems. Nevertheless, a study of the impact on the quality of the ES-MDA itself should be realized as well.

\section*{\textbf{Acknowledgments}}

The authors would like to thank Petr\'oleo Brasileiro S.A. (PETROBRAS) for funding this work,  Computer Modeling Group (CMG) for providing the simulator used in this research, and Laborat\'{o}rio Nacional de Computa\c{c}\~{a}o Cient\'{i}fica (LNCC) for providing their HPC infrastructure for the experiments.

\section*{\textbf{Conflicts of interest/Competing interests}}

All authors certify that they have no affiliations with or involvement in any organization or entity with any financial or non-financial interest in the subject matter or materials discussed in this manuscript.

\section*{\textbf{Author's contributions}}

\textbf{Felipe Portella} conceived the initial idea/methodology, wrote the algorithm, carried out the experiments, and wrote the manuscript as a Ph.D. student under the supervision of \textbf{Josep Ll. Berral}. \textbf{David Buchaca} helped to conceive the methodology, provided helpful support in the machine learning parts of the algorithm, and in review and editing of the manuscript. \textbf{José Roberto} provided valuable discussions about the algorithm strategy, and the results, along with helpful reviews of the manuscript. The authors read and approved the final manuscript.

\bibliographystyle{elsarticle-num}
\bibliography{library.bib}

\newpage

\authorbibliography[scale=0.40,wraplines=9,overhang=40pt,imagewidth=4cm,imagepos=r]{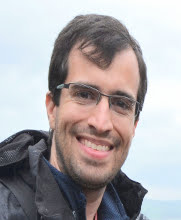}{Felipe Albuquerque Portella}{He is an IT Consultant at the Brazilian energy company Petr\'oleo Brasileiro S.A. (PETROBRAS), working with petroleum reservoir simulation workloads in HPC environments. He received his degree in Informatics (2003) and an M.Sc in Computer Science (2008) at PUC-Rio. He is currently a Ph.D. student at the BarcelonaTech-UPC in partnership with the Barcelona Supercomputing Center (BSC-CNS).}

\authorbibliography[scale=0.15,wraplines=8,overhang=40pt,imagewidth=4cm,imagepos=r]{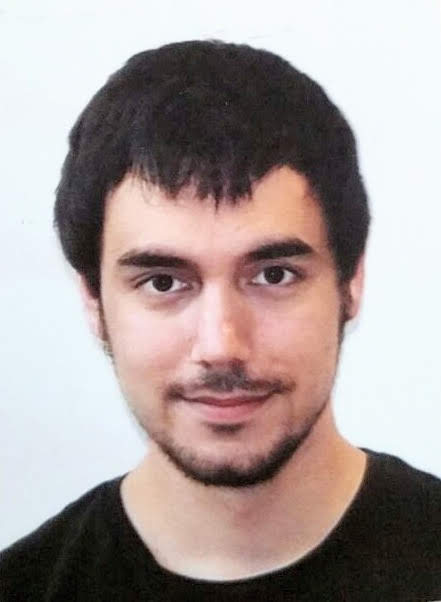}{David Buchaca Prats}{He received his degree in Mathematics (2012), M.Sc in Artificial Intelligence (2014), and Ph.D. at BarcelonaTech-UPC (2021). He is an applied mathematician, working in applications of Artificial Neural Networks, currently developing neural information retrieval systems.}

\authorbibliography[scale=0.3,wraplines=8,overhang=40pt,imagewidth=4cm,imagepos=r]{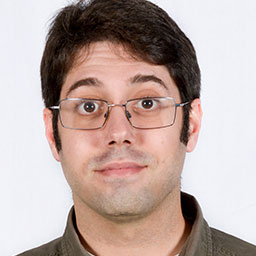}{Josep Llu\'is Berral Garcia}{He received his degree in Informatics (2007), M.Sc in Computer Architecture (2008), and Ph.D. at BarcelonaTech-UPC (2013). He is a data scientist, working in applications of data mining and machine learning on data-center and cloud environments at the Barcelona Supercomputing Center (BSC) within the ``Data-Centric Computing'' research line. In 2017, he received a Juan de la Cierva research fellowship from the Spanish Ministry of Economy. He is an IEEE and ACM member.}

\authorbibliography[scale=0.4,wraplines=10,overhang=40pt,imagewidth=4cm,imagepos=r]{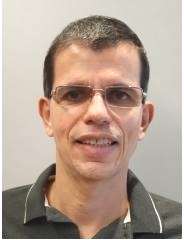}{Jos\'e Roberto Pereira Rodrigues}{He is a senior researcher in the reservoir engineering team at the Petrobras R\&D Center (CENPES). He has worked on several aspects of reservoir simulation, including linear and nonlinear solvers, inverse problems, optimization, and modeling of lab experiments. He holds B.Sc, M.Sc, and Ph.D. degrees in Mathematics from the Federal University of Rio de Janeiro.}

\end{document}